\documentclass[10pt,superscriptaddress,tightenlines,twocolumn,amsmath,amssymb,aps, pra]{revtex4-2}

\usepackage[utf8]{inputenc}
\usepackage{CJKutf8}

\UseRawInputEncoding
\usepackage{graphicx}
\usepackage{dcolumn}
\usepackage{bm}
\usepackage{bbm}
\usepackage{color}
\usepackage{amsmath}
\usepackage{amssymb}
\usepackage{amsthm}
\usepackage{subfigure}
\usepackage{braket}
\usepackage{indentfirst}
\usepackage{mathrsfs}
\usepackage[linesnumbered,ruled,vlined]{algorithm2e}
\usepackage{enumerate}
\usepackage{enumitem}
\usepackage{algorithmic}
\usepackage{amsfonts}
\usepackage[normalem]{ulem} %package for delete

\usepackage{makecell}
\usepackage{lmodern}
\usepackage{lineno}
\usepackage{multirow}

\usepackage{physics}
\usepackage{braket}
\usepackage{hyperref}

\newcommand{\be}{\begin{equation}}
\newcommand{\ee}{\end{equation}}

\newcommand{\RNum}[1]{\uppercase\expandafter{\romannumeral #1\relax}}

\def \ket#1{|#1\rangle}

\makeatletter

\newcommand{\Rmnum}[1]{\expandafter\@slowromancap\romannumeral #1@}

\makeatother

\usepackage{xcolor}
\colorlet{RED}{red}
\colorlet{BLACK}{black}

\providecommand{\U}[1]{\protect\rule{.1in}{.1in}}

\hypersetup{colorlinks=true, linkcolor=red, citecolor=blue, urlcolor=black}
\makeatletter
\newcommand{\newreptheorem}[2]{\newtheorem*{rep@#1}{\rep@title}\newenvironment{rep#1}[1]{\def\rep@title{#2 \ref*{##1}}\begin{rep@#1}}{\end{rep@#1}}}
\makeatother

\usepackage{amsthm}

\newtheorem{theorem}{Theorem}
\newreptheorem{theorem}{Theorem}

\newreptheorem{lemma}{Lemma}

\newreptheorem{proposition}{Proposition}
\newtheorem{definition}{Definition}

\begin{document}
\title{Experimental asymmetric relativistic zero-knowledge proofs with unconditional security}

\author{Chen-Xun Weng}
\affiliation{National Laboratory of Solid State Microstructures and School of Physics, Collaborative Innovation Center of Advanced Microstructures, Nanjing University, Nanjing 210093, China}
\affiliation{School of Physics and Beijing Key Laboratory of Opto-electronic Functional Materials and Micro-nano Devices, Key Laboratory of Quantum State Construction and Manipulation (Ministry  of  Education), Renmin University of China, Beijing 100872, China}
\affiliation{Centre for Quantum Technologies, National University of Singapore, Singapore 117543, Singapore}
\author{Ming-Yang Li}
\affiliation{National Laboratory of Solid State Microstructures and School of Physics, Collaborative Innovation Center of Advanced Microstructures, Nanjing University, Nanjing 210093, China}
\affiliation{School of Physics and Beijing Key Laboratory of Opto-electronic Functional Materials and Micro-nano Devices, Key Laboratory of Quantum State Construction and Manipulation (Ministry  of  Education), Renmin University of China, Beijing 100872, China}
\author{Nai-Rui Xu}
\affiliation{MatricTime Digital Technology Co. Ltd., Nanjing 211899, China}
\author{Yanglin Hu}
\author{Ian George}
\author{Jiawei Wu}
\affiliation{Centre for Quantum Technologies, National University of Singapore, Singapore 117543, Singapore}
\author{Shengjun Wu}\email{sjwu@nju.edu.cn}
\affiliation{National Laboratory of Solid State Microstructures and School of Physics, Collaborative Innovation Center of Advanced Microstructures, Nanjing University, Nanjing 210093, China}
\author{Hua-Lei Yin}\email{hlyin@ruc.edu.cn}
\affiliation{School of Physics and Beijing Key Laboratory of Opto-electronic Functional Materials and Micro-nano Devices, Key Laboratory of Quantum State Construction and Manipulation (Ministry  of  Education), Renmin University of China, Beijing 100872, China}
\affiliation{National Laboratory of Solid State Microstructures and School of Physics, Collaborative Innovation Center of Advanced Microstructures, Nanjing University, Nanjing 210093, China}
\author{Zeng-Bing Chen}\email{zbchen@nju.edu.cn}
\affiliation{National Laboratory of Solid State Microstructures and School of Physics, Collaborative Innovation Center of Advanced Microstructures, Nanjing University, Nanjing 210093, China}
\affiliation{MatricTime Digital Technology Co. Ltd., Nanjing 211899, China}

\date{\today}

\begin{abstract}
Zero-knowledge proofs (ZKPs) are widely applied in digital economies, such as cryptocurrencies and smart contracts, for establishing trust and privacy between untrusted parties. Classical ZKPs rely on computational assumptions and are vulnerable to quantum attacks. While a recent advance suggests quantum-sound symmetric relativistic ZKPs for the graph three-coloring problem without computational assumptions, the high round complexity, which leads to unachievable runtime and overall randomness cost, renders them impractical for real-life deployment. To overcome this, we develop an efficient asymmetric relativistic ZKP protocol using relativistic bit commitments, and prove its quantum soundness by relating it to the nonlocal Clauser-Horne-Shimony-Holt (CHSH) game. Our protocol achieves a linear relationship between the round complexity and the number of edges, and thus significantly improves practical feasibility. In addition, we implement a proof-of-principle experiment which completes all interactive rounds in about 0.22 seconds and requires an overall randomness cost of 430.81 MB. Our work illustrates the powerful potential of integrating special relativity with quantum theory in trustless cryptography, paving the way for robust applications against quantum attacks in distrustful Internet environments.

\end{abstract}

\maketitle

%%%%%%%%%%%%%%%%%%%%%%%%%%  body  %%%%%%%%%%%%%%%%%%%%%%%%%%
%\section{Introduction}

With the rapid advancement of information technology, the internet, particularly mobile internet, has brought significant convenience to everyone, enabling many important activities to be conducted online. In online activities, almost everyone is required to provide sensitive personal information, including facial data and personal fingerprint. However, personal privacy information is being arbitrarily exploited online, posing significant threats to both societal and personal
security. To address these risks, it is essential to consider how to perform tasks in an untrusted environment without disclosing personal privacy information, which is precisely the role of cryptographic primitives such as zero-knowledge proofs (ZKPs)~\cite{goldreich2001foundations}. In the middle of the 1980s, ZKP~\cite{goldwasser1985knowledge} was proposed to enable the prover to convince the verifier that a statement is true, while the verifier cannot learn any useful extra information of this statement. This system is particularly useful in scenarios where privacy and security are paramount, such as blockchain transactions, identity verification, secure communication protocols, and even nuclear warhead verification~\cite{chazelle2007security,glaser2014zero,philippe2016physical,chazelle2007security,glaser2014zero}.

Classical ZKP protocols rely on computational assumptions, such as the existence of one-way functions, raising concerns about their long-term security. This issue has become more pressing with advancements in quantum computing~\cite{shor1994algorithms,shor1999polynomial,o2007optical,arute2019quantum,fedorov2018quantum}, highlighting the need to address this vulnerability. Quantum zero-knowledge proofs, which involve the transmission of quantum information during interactions, offer a potential alternative to the computational assumptions of classical ZKPs. However, quantum rewinding in quantum ZKPs presents significant challenges due to the no-cloning theorem~\cite{watrous2006zeroknowledge}, making its implementation difficult. Additionally, unconditionally secure quantum bit commitment schemes are impossible~\cite{lo1997isquantum,mayers1997unconditionally}, and current quantum bit commitment schemes still rely on computational assumptions such as quantum one-way functions~\cite{yan2015quantum,kashefi2007statistical,ji2018pseudorandom,ananth2022cryptography}, which render it infeasible to directly replace classical bit commitment in some ZKP protocols to achieve unconditional security. 

Relativistic ZKPs (RZKPs) have been proposed and have attracted attention since they replace computational assumptions with special relativity, and the relativistic multi-prover interactive proof systems have been proven to be able to achieve quantum zero-knowledge without quantum rewinding~\cite{chailloux2017relativistic,crepeau2019practical,chailloux2021relativistic,crepeau2023zeroknowledge,shi2024relativistic}. In a recent breakthrough, Alikhani \textit{et al.} achieved the first experimental realization of a symmetric RZKP protocol against classical provers for the graph three-coloring problem without relying on computational assumptions~\cite{alikhani2021experimental}. It indicates that RZKPs have significant potential for widespread application in financial activities, allowing secure transactions without the need to disclose sensitive information~\cite{brassard2021relativity}.

However, in the quantum setting, establishing quantum soundness of RZKPs is non-trivial. Quantum entanglement between malicious provers can affect soundness analysis~\cite{scott2008thepower}. The classical-sound symmetric RZKP protocol that was experimentally demonstrated in~\cite{alikhani2021experimental} is vulnerable to such quantum attacks, as quantum entanglement may allow cheating provers to generate the required correlations and thus pass the verification even when the graph is not three-colorable~\cite{brassard2021relativity}. To address this, refs.~\cite{alikhani2021experimental,crepeau2019practical} propose a potential approach which introduces a third prover-verifier pair and increases the number of interactive rounds to achieve quantum soundness. However, this quantum-sound symmetric RZKP protocol suffers from extremely low data efficiency and high round complexity of $\mathcal{O}(|E|^4)$ ($|E|$: the number of edges) to ensure quantum soundness, making it impractical for graphs of realistic size. The inefficiency arises because, in each round, the verifiers must choose between two tasks: checking that connected nodes have different colors (proof check) or verifying that the provers share the same secret for the graph without lying (consistency check). Since these two checks cannot be performed simultaneously, many additional interactive rounds are required.

To achieve quantum soundness with reasonably low round complexity, inspired by the classical Goldreich-Micali-Wigderson (GMW) ZKP~\cite{Goldreich1991proofs}, we develop a relativistic variant that eliminates computational assumptions by replacing single-prover classical bit commitments with two-prover relativistic bit commitments~\cite{chailloux2017relativistic}, whose security relies on the no-signaling principle~\cite{kent1999unconditionally,lunghi2015practical,chakraborty2015arbitrarily,Verbanis201624-Hour,Chakraborty2016Robust}. Our main contributions are threefold: (a) We design an asymmetric relativistic GMW ZKP protocol that reduces the round complexity from $\mathcal{O}(|E|^4)$ to $\mathcal{O}(|E|)$ while also lowering the overall randomness cost compared with the quantum-sound symmetric protocol in~\cite{alikhani2021experimental,crepeau2019practical}. (b) We show that integrating two-prover relativistic bit commitments into the GMW framework is non-trivial in the quantum soundness analysis, as quantum-correlated provers can still alter their committed values with non-zero probability. We relate this scenario to the nonlocal CHSH$_Q(2)$ game and provide a rigorous upper bound for the quantum soundness error. (c) We implement a proof-of-principle experiment of our asymmetric RZKP over a $300$-meter separation for a three-colorable graph with $588$ vertices and $1097$ edges, completing all required interactive rounds in $0.22$ seconds. In contrast, the quantum-sound protocol in~\cite{alikhani2021experimental} would require approximately $10^{4}$ years. This highlights the practical efficiency of our asymmetric RZKP approach in real-life scenarios. In addition, we also discussed the scalability challenges of the practical implementation of asymmetric RZKPs in the real-life fiber-optical networks and the possible solutions to these scalability challenges.

\bigskip
\noindent
\textbf{\large Results}\\ 
Graph three-coloring problem is a nondeterministic polynomial-time (NP) complete problem, implying that any problem within the complexity class NP can be reduced to a graph three-coloring problem. Formally, a graph, $\mathbb{G}(V, E)$, where $V$ is the set of vertices and $E$ is the set of edges, is three-colorable if its vertices can be colored with only three different colors (denoted as $\forall k\in V, y_k \in \mathbb{F}_3$) such that no two adjacent vertices share the same color, i.e., $\forall \{u, v\}\in E$, $y_u\neq y_v$. (See Supplementary Note 1.2 of the Supplementary Information for details.)

Our asymmetric relativistic GMW ZKP protocol for graph three-coloring involves two verifier-prover pairs. The two provers, P1 and P2, aim to convince the verifiers that the graph is three-colorable, without revealing any useful information about the actual coloring.

\begin{figure}[t]
	\centering
        \includegraphics[width=8.5cm]{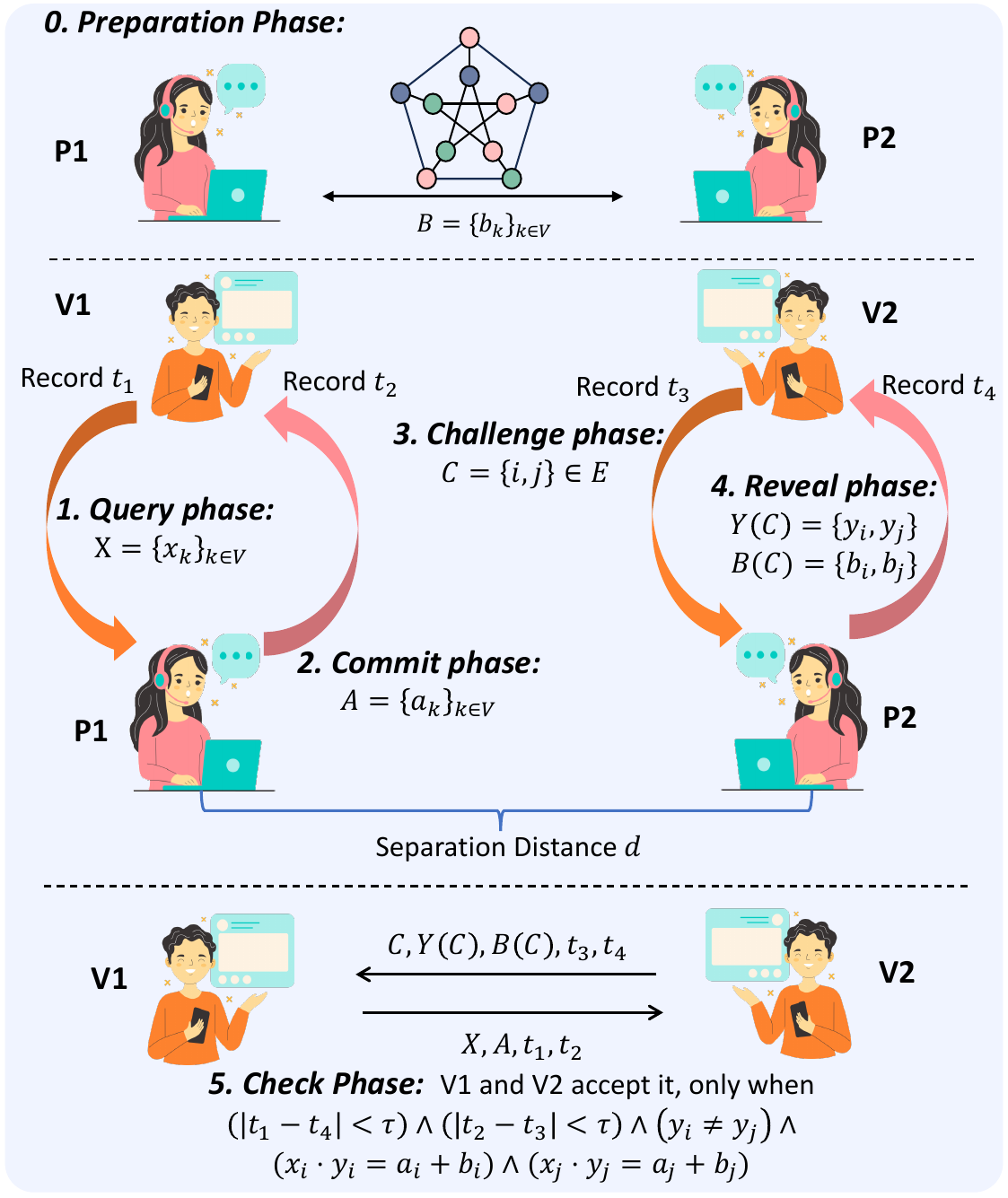}
\caption{\textbf{Schematic of asymmetric RZKP.} Two separate provers aim to convince verifiers that they possess knowledge of valid three-coloring for the graph, without revealing any information that could be used to reconstruct the three-coloring.  }

\label{fig1}
\end{figure}

\bigskip
\noindent
\textbf{Protocol description}\\
The protocol consists of $m$ rounds of interaction. As illustrated in Fig.~\ref{fig1}, we take one interactive round as an example. Note that all calculations are performed over the finite field $\mathbb{F}_Q$.

0. Preparation phase: P1 and P2 agree in advance on a uniformly random color permutation, $\pi$, for the three-colorable graph, where $\pi\in \Pi$ and $\Pi$ is the set of all valid permutations of three-colorings. Additionally, P1 and P2 prepare a set of uniformly random numbers $B = \{b_k \}_{k \in V}$, where $b_k \in \mathbb{F}_Q$, with $Q$ denoting the size of the finite field used in the protocol. Note that in the preparation phase of each round, the provers must select a fresh random color permutation $\pi$ and new random values for the set $B$ to protect the zero-knowledge property. (See ``Shared randomness'' of Methods for details).
    
1. Query phase: At time $t_1$, V1 generates a set of uniformly random numbers $X = \{x_k \}_{k \in V}$, where $x_k \in \mathbb{F}_Q$ and $x_k \neq 0$, and sends them to P1 as queries to request commitments to the colors of all vertices in the graph.

2. Commit phase: Upon receiving $X$, P1 immediately replies V1 with $A=\{a_k\}_{k \in V}$, where $a_k=x_k\cdot y_k-b_k$ and $y_k\in \mathbb{F}_3$ is the color of vertex $k$. V1 records the time of receiving the commitments from P1, denoted as $t_2$. 

3. Challenge phase: At time $t_3$, V2 chooses a uniformly random edge as a challenge, $C=\{i, j\}\in E$, and sends it to P2. 
    
4. Reveal phase: P2 announces the colors $Y(C)=\{y_i,y_j\}$ and sends its corresponding $B(C)=\{b_i, b_j\}$ to V2 for revealing the colors of $i$ and $j$. V2 records the time of receiving $B(C)$ from P2, denoted as $t_4$.
    
5. Check phase: (a) V1 and V2 first verify that $|t_1-t_4|<\tau$ and $|t_2-t_3|<\tau$, where $\tau=d/c$ is the reliable time separation between P1 and P2, with $d$ denoting the distance between P1 and P2, and $c$ the speed of light in vacuum; (b) V1 and V2 then check the following conditions: (1) $y_i\neq y_j$ and $y_{i},y_j\in \mathbb{F}_3$ (proof check), and (2) $x_i\cdot y_i=a_i+b_i$ and  $x_j\cdot y_j=a_j+b_j$ (consistency check). The protocol is accepted if all conditions are satisfied. 

If V1 and V2 accept all $m$ interactive rounds, they are convinced that the statement is true in which the graph is indeed three-colorable.

\begin{figure}[t]
	\centering
        \includegraphics[width=0.45\textwidth]{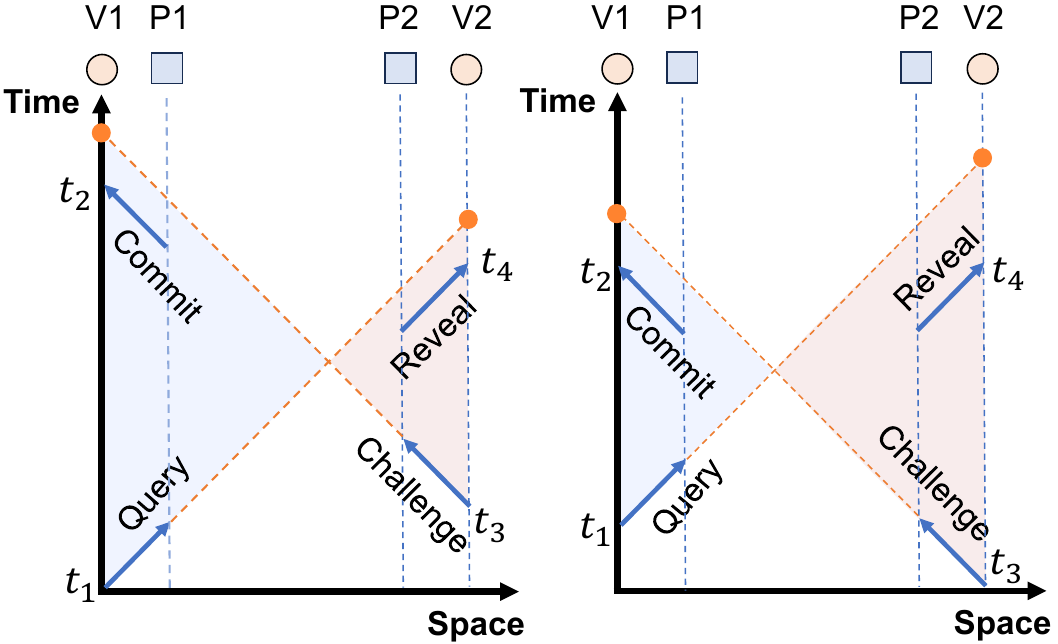}
\caption{\textbf{Space-time diagram.} The orange dashed diagonal line represents the relativistic constraints for the spatial configuration (the speed of light). V1 sends queries to P1 who commits to the colors. V2 sends a random challenge to P2, who reveals the commitments according to the challenge. The query and challenge phases do not need to follow a fixed time order.} 

\label{fig2}
\end{figure}

Figure~\ref{fig2} illustrates that the no-signaling principle ensures P2 has no information about the query $X$ before P2's response reaches V1 if $|t_1-t_4|<\tau$, and P1 has no information about the random challenge $C$ before P2's response reaches V2 if $|t_2-t_3|<\tau$.

\bigskip
\noindent\textbf{Quantum soundness and round complexity}\\
In the classical GMW ZKP for graph three-coloring with one prover and one verifier, the strong binding property of the bit commitment scheme plays a crucial role in the soundness analysis~\cite{goldwasser1985knowledge}. Suppose the graph is not three-colorable and contains one edge whose vertices are assigned the same color, say color 2. Due to the strong binding property, the probability that the prover can successfully open the commitment to 2 is $p_2 = 1$ and it is impossible to change the committed color to 0 or 1. Consequently, in a single round, the verifier catches the malicious prover with probability at least $\frac{1}{|E|}$, and the soundness error is $\delta=1 - \frac{1}{|E|}$. Thus, to achieve the overall soundness error of $\delta_s=\delta^m=e^{-k}$, classical GMW protocol requires at least $m=k|E|$ rounds. Note that such a strong binding is achievable only under computational assumptions.

However, in the case of two malicious quantum correlated provers, the classical soundness analysis no longer applies for our relativistic GMW ZKP. At first glance, cheating strategies of quantum correlated provers seem infeasible due to the binding property of relativistic bit commitment, similar to the classical protocol. Nevertheless, relativistic bit commitments can only guarantee a weaker notion known as sum-binding, which could only ensure that $p_0 + p_1 + p_2 \leq 1 + \varepsilon_b$~\cite{lunghi2015practical}. For example, a distribution such as $p_0 = \frac{1}{3} + \varepsilon_b$, $p_1 = \frac{1}{3}$, and $p_2 \leq \frac{1}{3}$ satisfies the sum-binding condition. In this case, if the verifier selects an invalid edge (i.e., one with two identical colors such as color 2), the malicious quantum provers may still be able to successfully open the commitment to a different color, since the sum $p_0 + p_1$ is not guaranteed to be small. That is why sum-binding does not offer composable security and must be carefully analyzed in each composite cryptographic construction~\cite{kaniewski2013secure}.

Fortunately, the quantum soundness of our protocol against quantum-correlated provers can be established without concretely analyzing the sum-binding property of relativistic bit commitment. Instead, when analyzing the soundness error of a single round in our protocol, it can be directly analyzed by viewing the successful cheating as a non-local game involving P1 and P2, when the invalid edge is challenged by the verifiers. Here we give the formal definition of quantum soundness in a quantum interactive proof system for NP problems~\cite{watrous2006zeroknowledge,regev2006quantum,Unruh2012quantum} (See Methods for the definitions of completeness and quantum zero-knowledge).

\begin{definition}[Quantum Soundness]
An interactive proof system $(P, V)$ for an NP relation $R$ has quantum soundness error $\delta_s$ if, for every malicious quantum prover $P^*$, every auxiliary quantum state $\ket{\Phi}$, and every instance $x \in L_{\mathrm{no}}$ such that for every classical witness $w$ with $(x,w)\notin R$,
\[
\Pr\!\left[ \big\langle P^*\!\left((x,w),\ket{\Phi}\right),\, V(x) \big\rangle = 1 \right]
\le \delta_s(|x|).
\]
\end{definition}

The task of the quantum soundness analysis is to show that for any input instance $x \in L_{\mathrm{no}}$ (every witness candidate is invalid) and for arbitrary classical witness candidate $w$ given to quantum provers, any cheating strategy of provers involving quantum operations or shared entanglement allows such arbitrary $w$ to pass the verifier's checks only with negligible probability~\cite{watrous2006zeroknowledge,Unruh2012quantum,chailloux2017relativistic,crepeau2023zeroknowledge}.

In this work, since $x \in L_{\mathrm{no}}$ (the graph is not three-colorable), every such witness $w$ (a coloring assignment) necessarily contains at least one invalid edge with identical colors in the two vertices. During the execution of the RZKP protocol, the quantum provers
may share arbitrary entangled auxiliary states and apply quantum
strategies throughout the commitment and reveal phases. Nevertheless, by the
definition of quantum proof systems for NP problems, the witness itself remains classical in the graph three-coloring problem (see
Methods for NP definitions). Quantum soundness error therefore upper-bounds the probability that quantum provers can cause an arbitrary invalid coloring to be accepted. Note that RZKP protocol for NP is not intended to address more general adversaries that are given quantum witnesses or settings where no classical witness exists (in such cases, the protocol would no longer be a proof for an NP relation, see Methods).

We therefore consider a general setting in which, at the beginning of each round, the provers are given an arbitrary classical witness $w=\{y_i\}_{i\in V}$ (any possible color assignment). Because the the graph is not three-colorable, every possible classical witness can be partitioned into two disjoint edge sets, $E=E_{\text{c}}\cup E_{\text{inc}}$, where $E_{\text{c}}$ consists of edges whose vertices receive different colors, and $E_{\text{inc}}$ consists of edges incorrectly colored with the same color.  If the verifier's random challenge selects an edge in $E_{\text{c}}$, the provers always succeed without changing their committed colors. However, when an edge in $E_{\text{inc}}$ is selected, malicious provers must apply quantum or classical cheating strategies to alter the committed colors during the commit and reveal phase. We show that the probability of successfully cheating in this case is upper-bounded by the quantum value of the nonlocal CHSH$_Q(2)$ game due to the no-signaling principle. Consequently, the quantum soundness error of our asymmetric RZKP can be obtained as follows (see Methods for the full proof of quantum soundness).

\begin{theorem}[Quantum soundness of asymmetric RZKP for NP]\label{quantum_soundness}
In a single round of our asymmetric RZKP protocol for the graph three-coloring problem, if the graph is not three-colorable ($x\in L_{no}$), consider an arbitrary classical witness $w$ (a color assignment) given to quantum provers. For any cheating strategy of quantum provers involving arbitrary quantum operations and shared entanglement during the commitment and reveal phase, the verifier's acceptance probability (i.e., quantum soundness error) is upper bounded by
\begin{equation}\label{soundness_1}
\delta \leq 1 - \frac{1}{|E|}+\frac{1}{|E|}\omega^*(\text{\rm CHSH}_Q(2)),
\end{equation}
where $\omega^*(\text{\rm CHSH}_Q(2))\leq \frac{1}{2}+\frac{1}{\sqrt{2Q}}$~\cite{sikora2014strong,shi2024relativistic,weng2025tight}. Accordingly, after $m$ rounds of interaction, the overall quantum soundness error satisfies
\begin{equation}\label{soundness_2}
\delta_s = \delta^m \leq \left(1- \frac{1}{2|E|}+\frac{1}{\sqrt{2Q}|E|} \right)^m.
\end{equation}
In the regime where both $Q$ and $|E|$ are large, choosing $Q = |E|^2$ to ensure that $\sqrt{2Q}|E|\gg 2|E|$, which implies a per-vertex randomness cost of $N=\log Q = 2\log |E|$ bits, guarantees that achieving an overall quantum soundness error of $\delta_s=e^{-k}$ requires at least
\begin{equation}
m = 2k|E|.
\end{equation}
\end{theorem}

We define the total number of interactive rounds required to achieve an overall soundness error of $\delta_s = e^{-k}$ as the round complexity. The key difference of soundness and round complexity between relativistic and classical GMW protocol lies in the fact that quantum-correlated provers could change their committed color with non-ignored probability when the challenge is an invalid edge.

\bigskip
\noindent\textbf{Randomness cost}\\
In ZKP protocols, it is important to quantify the amount of randomness required. Here, we define the randomness cost of a single round as the total number of random bits used by all parties, including both provers and verifiers. In our protocol, this randomness cost can be expressed as
\begin{equation}\label{RC}
\begin{aligned}
\text{RC} &= \text{RC(P1 and P2)} + \text{RC(V1 and V2)} \\
&= |V|\log Q + (|V|\log Q + \log |E|) \\
%&= 2|V|\log |E| + \log |E| \\
&= (4|V| + 1)\log |E|,
\end{aligned}
\end{equation}
where $|V|\log Q$ random bits are used by V1 in the query phase to generate $X = \{x_k\}_{k \in V}$, another $\log |E|$ random bits are used by V2 in the challenge phase to select a random edge $C $$=$$ \{i, j\} $$\in$$ E$, and additional $|V|\log Q$ random bits, corresponding to $B = \{b_k\}_{k \in V}$, are pre-shared between P1 and P2 for generating the commitments. Since we set $Q = |E|^2$ to ensure soundness, the total randomness cost is simplified to $(4|V| + 1)\log |E|$ per round. In addition, we can obtain the overall randomness cost of our protocol as
\begin{equation}\label{overall_RC}
    \text{Overall RC}=\text{RC}\times m=2k(4|V|+1)|E|\log|E|,
\end{equation}
where $m$ is the round complexity of our protocol to achieve the overall soundness error of $e^{-k}$.

For a fair comparison, we also evaluate the randomness cost of the quantum-sound protocol in~\cite{alikhani2021experimental}. In their scheme, which involves three prover-verifier pairs, the per-round randomness cost is $\text{RC} = 3(1 + \log |E|)$. Specifically, in the P1-V1, P2-V2, and P3-V3 pairs, each verifier (V1, V2, or V3) sends a random edge together with a random bit to the corresponding prover, while the provers themselves require no randomness. However, when accounting for the extremely high round complexity of their quantum-sound protocol, $k(11|E|)^4$, the overall randomness cost scales as $11^4 k|E|^4 (3 + 3\log |E|)$.

\begin{figure*}[t]
	\centering
        \includegraphics[width=17cm]{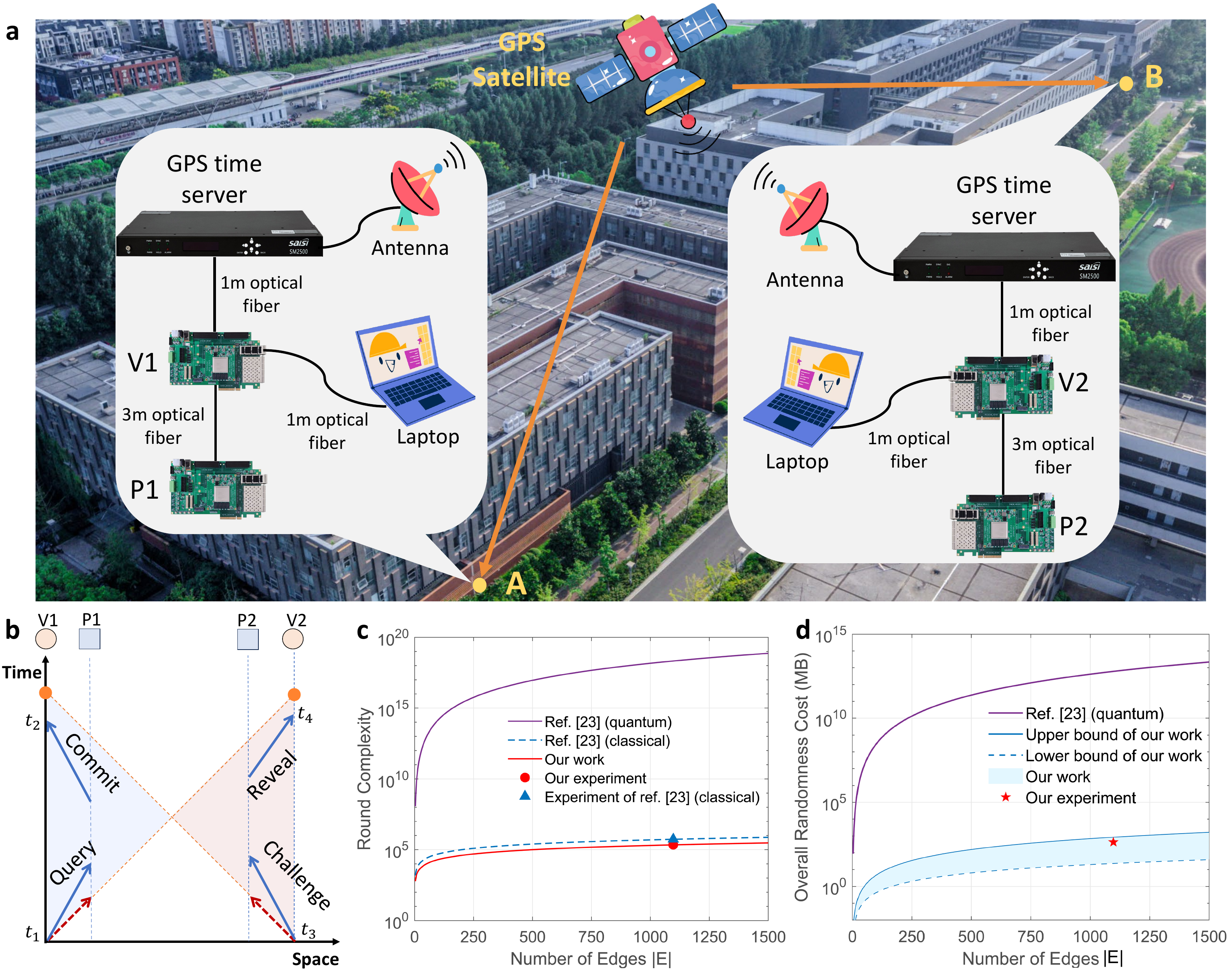}
	\caption{\textbf{Experimental implementation.} \textbf{a} Satellite view. The distance between the two P-V pairs is 300 meters. This separation makes the direct communication between P1 and P2 impossible within 1000 ns due to the no-signaling principle.  \textbf{b} Space-time diagram of the experiment. In the experiment, the query and challenge phase are set to be triggered simultaneously. The solid blue lines represent the real signal with delays including electrical delays, hardware response delays and so on.  The diagonal red dashed lines represent the ideal signal at the speed of light. \textbf{c} Round complexity. We compare the round complexity between our work and \cite{alikhani2021experimental}. In \cite{alikhani2021experimental}, the round complexity of classical-secure protocol is $5k|E|$ while that of quantum-secure protocol is increased to $k(11|E|)^4$. In our work against malicious quantum-correlated provers, the round complexity is $2k|E|$.  \textbf{d} Overall randomness cost.  We compare the overall randomness cost of our work and \cite{alikhani2021experimental} to achieve quantum soundness error of $e^{-100}$. In the experiment, for the same graph with $|V|$$=$$588$ and $|E|$$=$$1097$, the quantum-sound symmetric protocol of \cite{alikhani2021experimental} requires about $5.92\times10^{9}$ GB random bits while our quantum-sound asymmetric RZKP protocol requires only 430.81 MB.} 
\label{fig3}
\end{figure*}

\bigskip
\noindent
\textbf{Experimental implementation} \\
We now describe the experimental implementation of our RZKP. Our protocol involves two separated verifier-prover pairs. Figure~\ref{fig3}a shows the details of our experiment. On each prover's side, the prover is equipped with a field-programmable gate-array (FPGA) card (Xilinx Ultrascale FMC HPC board featuring MLK-H5-KU040/060) to reduce communication latency, speed up computation, and improve time reliability. Each verifier has a computer (Intel core i9 processor with 32 GB RAM) and an FPGA card to handle the computations and record the time and messages. To ensure accurate time synchronization between V1 and V2, their FPGAs are both synchronized with Coordinated Universal Time (UTC) via the Global Positioning System (GPS).

\begin{table}[b]
\caption{\textbf{Experimental data.} $|t_1-
t_4|$ and $|t_2-t_3|$ are all below 1000 nanoseconds (ns). The table presents the mean, maximum (Max), minimum (Min), median, and standard deviation (SD) of the time differences. ns: nanosecond. The details of experimental data can be found in the supplementary files.}
\resizebox{\linewidth}{!}{ 
\begin{tabular}{lccccc}
\hline\hline
 &Max(ns)  &  Min(ns) &   Mean(ns)&  Median(ns) & SD(ns)\\ \hline
$|t_1$$-$$t_4|$ & 879.91 & 505.97 & 712.87
&712.06 & 15.63\\
$|t_2$$-$$t_3|$ &876.77  &510.88 &684.61 &684.34 &15.98 \\
\hline\hline
\end{tabular}}\label{table1}
\end{table}

\begin{table*}[t]
\caption{\textbf{The comparison of ZKPs for the graph three-coloring problem.} $k$ is the security parameter of overall soundness error which is $e^{-k}$. $\varepsilon_h$ is the security parameter of the hiding of QCRHF. BC: bit commitment. PQC: post-quantum cryptography. QCRHF: quantum collision-resistant hash function. QOWF: quantum one-way function. QOWP: quantum one-way permutation. N/A: not applicable. For any graph $\mathbb{G}(V, E)$, the number of vertices and edges satisfies $|V| \leq |E| \leq \frac{|V|(|V| - 1)}{2}$. Therefore, $|V|$ is approximately $\mathcal{O}(|E|^{\beta})$ for some $\beta \in [1/2, 1]$. }\label{table2}
\resizebox{\linewidth}{!}{ 
\begin{tabular}{lccccc}
\hline\hline
 & \textbf{Our work }  &  \textbf{Ref.~\cite{alikhani2021experimental} }          &  \textbf{Ref.~\cite{crepeau2019practical,alikhani2021experimental}}      &   \textbf{Ref.~\cite{li2023device} } &   \textbf{Ref.~\cite{watrous2006zeroknowledge} }   \\ \hline
Cryptography tool & relativistic BC  & special relativity & special relativity & \makecell[c]{QCRHF~$\&$~PQC}  %\\PQC signature 
& QOWF (QOWP)\\
No computational assumptions&$\checkmark$    &$\checkmark$  &$\checkmark$  &$\times$ &$\times$\\
Quantum soundness          & $\checkmark$ & $\times$      &     $\checkmark$    & $\times$ & $\checkmark$ \\ 
Quantum zero-knowledge         & $\checkmark$ & $\checkmark$    &     $\checkmark$    & $\times$   & $\checkmark$  \\ 
Number of provers           & 2 &2      & 3     &1   &1      \\ 
Round complexity & $2k|E|$ &  $5k|E|$   &  $k(11|E|)^4$ & $k|E|$  & $k|E|$\\ 
Randomness cost &  $(4|V|+1)\log|E|$
&  $2+2\log|E|$     & $3+3\log|E|$  & $(2$$-$$12\log\varepsilon_h)|V|$ &depend on QOWF\\ 
Overall randomness cost & $\mathcal{O}(|V||E|\log|E|)$ &  $\mathcal{O}(|E|\log|E|)$  &  $\mathcal{O}(|E|^4\log|E|)$  & $\mathcal{O}(|V||E|)$  & N/A\\ 
Message exchanged  &classical bits    &classical bits   &classical bits   &classical bits  &qubits\\
Experimental realization  &  $\checkmark$ &   $\checkmark$    & $\times$ & $\checkmark$ & $\times$
\\ \hline\hline
\end{tabular}}
\end{table*}

To maintain synchronization with the GPS reference, the 1-PPS (one pulse per second) signal from a high-precision GPS time synchronization server is sent directly to the verifiers' FPGA, and each FPGA checks synchronization based on this 1-PPS signal. The absolute time accuracy of the GPS time synchronization server is within $\Delta=30$ ns, including the delay from optical links between the server and FPGA. Each GPS time server is equipped with a rubidium atomic clock with better than 1 ns resolution, ensuring a time drift of less than 1 us/24h even if the GPS signal is lost.

The two pairs of verifier-prover are located separately in two buildings on the Xianlin campus of Nanjing University, 300 m apart, corresponding to a time separation of $\tau =1000$ ns. For the graph coloring, we pre-store a fixed three-colorable graph and all color permutations for this graph in the two provers' FPGAs. The graph used in the experiment has $|V| $$=$$588$ vertices and $|E| $$=$$ 1097$ edges, generated by the No-Choice algorithm~\cite{turner1988almost} (see Methods for details).

In each round, two provers' FPGAs prepare a permutation $\pi$ of coloring. V1's FPGA sends a random query $X$ to P1 (query phase), while V2's FPGA simultaneously sends a random challenge $C$ to P2 (challenge phase). P1 and P2 respond to the query and challenge, respectively. V1 records $t_1$ and $t_2$, while V2 records $t_3$ and $t_4$. V1 and V2 check whether the conditions $|t_1 - t_4| < \tau$ and $|t_2 - t_3| < \tau$ are satisfied. Table~\ref{table1} shows that both $|t_1 - t_4|$ and $|t_2 - t_3|$ for all rounds are consistently below $\tau = 1000$ ns. Considering the worst situation of the maximum accuracy error of GPS time server, $\Delta=30$ ns, on both two sides, the maximum of real $|t_1 - t_4|$ and $|t_2 - t_3|$ are $879.91+2\Delta=939.91$ ns and $876.77+2\Delta=930.77$ ns, which are also below $\tau = 1000$ ns. In our implementation, the interaction time between the provers and verifiers is primarily constrained by hardware latency. The computation of the provers' commitments can be completed within a single clock cycle of the FPGA (6.4 ns, with a working frequency of 156.25 MHz). However, the communication delay between the provers and verifiers, including the time for light signals in the fiber, light-electrical signal conversion, and hardware response times, is approximately 300 ns. Unlike previous works that applied unilateral time compensation for the commit and reveal phases~\cite{alikhani2021experimental}, which could introduce vulnerabilities allowing a prover to manipulate delay-time attacks and bypass relativistic spacetime constraints, we account for all communication and hardware delay errors without time compensation for the commit and reveal phases as shown in Fig.~\ref{fig3}b.

As shown in Fig.~\ref{fig3}c, to achieve an overall soundness error of $\delta_s=e^{-k}=e^{-100}$, our protocol requires approximately $m=2k|E|=2.2\times 10^{5}$ rounds, while quantum-sound symmetric RZKP protocol in~\cite{alikhani2021experimental} would require about $m=k(11|E|^4)=2\times10^{18}$ rounds, rendering it infeasible for real implementation. We set the trigger interval between rounds to 1 us in our experiment, allowing our protocol to be completed in roughly 0.22 seconds, whereas the quantum-sound RZKP of \cite{alikhani2021experimental} would take about $6.72\times 10^{4}$ years by applying the same trigger time interval.

For each commitment, according to Theorem~\ref{quantum_soundness}, the number of bits exchanged in each commitment is $N = 2\log|E|=22$. Thus, as illustrated in Fig.~\ref{fig3}d, the overall randomness costs over the experiment are  $430.81$ megabyte (MB) according to Eq.~\ref{overall_RC}, which is reasonable for practical daily use. However, in \cite{alikhani2021experimental}, it would require computational resources of $5.92\times10^{9}$ gigabytes (GB) to ensure the same quantum soundness because of its extremely high round complexity.

Additionally, the randomness in our ZKP can be enhanced with quantum random number generation~\cite{li2023device, miguel2017quantum}. For speed and simplicity, it is preferable to store this shared randomness in the FPGAs of the provers and verifiers. In each round of our experiment, the random color permutations $\pi$, random query $X$, random encoding keys $B$, and random challenges $C$ are pre-generated by a source-independent quantum random number generation~\cite{liu2023source}.

Finally, we present a detailed comparison of various relativistic and quantum ZKPs for graph three-coloring in Table~\ref{table2}. Although the graph three-coloring problem can be reduced to other NP-complete problems, such as the Hamiltonian cycle, via polynomial-time reductions, our comparison focuses on ZKPs for graph three-coloring. This is because such reductions are not unique, and constructing ZKPs for other problems through polynomial-time reduction may lead to substantial differences in round complexity and randomness cost. Consequently, ZKP protocols are often designed independently for different NP-complete problems to better optimize performance.

\bigskip
\noindent\textbf{Asymmetric and symmetric RZKPs}\\
Here, we summarize the advantages and limitations of our asymmetric relativistic GMW ZKP in comparison with the symmetric quantum-secure protocol of ref.~\cite{alikhani2021experimental,crepeau2019practical}. Our protocol is asymmetric, meaning that the V1-P1 pair is responsible for the majority of the randomness cost and computation load: V1 sends random bit strings for all vertices, and P1 encrypts the colors of all vertices, which amounts to $2|V|\log |E|$ bits, as shown in Eq.~\ref{RC}. By contrast, the quantum-sound protocol in ref.~\cite{alikhani2021experimental} is symmetric: the P1-V1, P2-V2 and P3-V3 pairs perform the same steps, each requiring only $\log |E| + 1$ bits of randomness per round. Therefore, this asymmetry introduces practical challenges for the V1-P1 setup in large graphs, potentially limiting the performance of our protocol when the graph contains a large number of vertices and edges.

Nevertheless, our protocol benefits from relativistic bit commitment in the asymmetric GMW framework, which enables both proof check and consistency check within a single round. This leads to a round complexity of $\mathcal{O}(|E|)$, allowing our protocol to be executed within a realistic time, something that is infeasible in that of ref.~\cite{alikhani2021experimental} for a large graph, which requires at least $\mathcal{O}(|E|^4)$ rounds. Moreover, as shown in Eq.~\ref{overall_RC}, in terms of the overall randomness cost, our protocol achieves $\mathcal{O}(|V||E|\log |E|)$, which is much lower than the $\mathcal{O}(|E|^4\log |E|)$ cost in ref.~\cite{alikhani2021experimental}. 

We compare the two protocols under the same security level of quantum soundness and observe that it is difficult to simultaneously minimize both the randomness cost per round and the round complexity. Our asymmetric protocol makes a trade-off by increasing the randomness cost per round due to the usage of relativistic bit commitments, thus leads to a significant decrease of the round complexity and the corresponding overall execution time of FPGAs. This brings the total run-time of quantum-secure RZKPs down to a practical and feasible level which is impossible for quantum-sound symmetric RZKP in~\cite{alikhani2021experimental}. Therefore, we note that achieving quantum soundness in RZKPs requires carefully balancing the trade-off between single-round randomness cost and round complexity.

\bigskip
\noindent	
\textbf{\large Discussion}\\
In conclusion, our asymmetric relativistic GMW ZKP relies only on commercially available hardware and uses relativistic bit commitments to resist quantum correlated provers to ensure quantum soundness. In our protocol, verifiers can do the proof check and consistency check simultaneously in a single round due to relativistic bit commitment, and thus our work reduces the round complexity from $\mathcal{O}(|E|^4)$ to $\mathcal{O}(|E|)$, significantly lowering runtime and overall randomness costs without adding hardware complexity, as in~\cite{alikhani2021experimental}. In our experiment, all interactive rounds are completed in approximately 0.22 seconds, with an overall randomness cost of 430.81 MB. In contrast, the quantum-sound symmetric RZKP in~\cite{crepeau2019practical,alikhani2021experimental} requires about $6.72 \times 10^{4}$ years of overall runtime and an overall randomness cost of $5.92 \times 10^{9}$ GB to achieve the same overall soundness error of $e^{-100}$, which is completely unattainable using everyday electronic devices. This highlights the practicality of our RZKP for real-world applications involving reasonably sized three-colorable graphs, even under malicious quantum correlated provers.

Due to its simplicity in hardware requirements, our protocol can even be implemented on everyday devices such as smartphones. Compared with quantum ZKPs~\cite{watrous2006zeroknowledge}, RZKPs transmit classical bits instead of qubits, require no computational assumptions, and rely solely on the no-signaling principle to guarantee quantum soundness and zero-knowledge. As quantum technology advances, our asymmetric RZKP offers strong potential for practical deployment in privacy-sensitive applications such as blockchains, smart contracts, anonymous e-voting, and online auctions. For example, in an anonymous electronic voting system, RZKPs can allow each voter to prove that their vote is valid without revealing their choice to anyone, including the verifier. This ensures both the integrity and the privacy of the election, while avoiding reliance on computational hardness assumptions that may be threatened by future quantum computers. Moreover, it is possible to extend our RZKP to a future topic called ZKPoK, which would enable unconditionally secure zero-knowledge identity verification. (See ``ZKP and ZKPoK'' of Methods for details.)

As shown in our comparison between symmetric and asymmetric RZKPs above, an interesting open question is whether a fundamental trade-off bound exists between round complexity and randomness cost per round in RZKPs against quantum provers, or whether it is possible to design a protocol that simultaneously achieves optimal performance on both sides. Moreover, the practical use of asymmetric RZKP in the real-life fiber-optical networks still faces several scalability challenges, such as the verification of accurate positions of two provers and computation load. (See ``Scalability of RZKPs'' of Methods for details.) In addition, another promising direction is to extend RZKPs for quantum Merlin-Arthur (QMA) problems~\cite{broadbent2020zero,coladangelo2020noninteractive,vidick2020classical}. Such extensions could yield practical ZKPs for QMA-complete problems without computational assumptions such as quantum one-way functions.

\bigskip
\noindent	
\textbf{\large Methods}\\	
\noindent		
\textbf{Three-colorable graph generation algorithm.}
For the experimental implementation, we require a concrete graph along with a corresponding three-coloring. While finding a three-coloring solution for an unfamiliar graph is difficult, several efficient algorithms exist to directly generate a three-colorable graph~\cite{mizuno2008constructive,turner1988almost}. In our experiment, to generate a three-colorable graph $\mathbb{G}(V, E)$, we use the No-Choice algorithm from ref.~\cite{turner1988almost}. The algorithm is described as follows: 

1. Randomly assign a color to each vertex, i.e., for each $u \in V$, let $y_u$ be a random integer in $\{0, 1, 2\}$. 

2. For each pair $\{u, v\} \in V\times V$ $(u\neq v)$ such that $y_u\neq y_v$, add an edge to the set $E$ with probability $p$, where $0<p<1$.

The value of $p$ affects the number of edges for a given $V$ and can even determine whether the three-coloring problem for the generated graph is hard. Therefore, in this work, $p$ should be chosen carefully, neither too large nor too small, to ensure that the graph is complex enough to find its three-colorability, without compromising the core idea of our ZKP protocol. 

\bigskip
\noindent		
\textbf{Shared randomness.}
In our RZKP, the two provers require access to shared randomness in each round, including random color permutations $\pi$ for the three-coloring and random values of the set $B=\{b_k\}_{k\in V}$. Since communication between the provers is disallowed during the protocol execution due to relativistic constraints, shared randomness must be established in the preparation phase. In our experimental implementation, we adopt a pre-storage strategy: the provers are initially co-located in a secure environment, where they agree on and store all the required randomness for all rounds before moving to their spatially separated locations. During protocol execution, the stored randomness is accessed sequentially, ensuring that each round uses a fresh color permutation and fresh random values for $B$. This per-round re-randomization is essential to maintain the zero-knowledge property against malicious verifiers. The pre-storage approach  offers both conceptual simplicity and experimental practicality.

Alternative methods commonly used in relativistic cryptographic protocols~\cite{liu2014experimental}, such as having one prover generate the randomness locally and transmit it securely to the other using encryption with quantum key distribution-generated secret keys~\cite{yin2016measurement,lucamarini2018overcoming,ma2018phase,xu2020secure,xie2022breaking}, or even direct transfer via a trusted carrier, can also ensure information-theoretic security.

\bigskip
\noindent	
\textbf{NP Languages and Witnesses.}
In our setting, the 3--coloring problem is treated strictly as an NP language. Here, we give the formal definition of NP problem~\cite{regev2006quantum}. The class NP consists of all languages \( L \subseteq \{0, 1\}^* \) for which there exists a uniformly generated family of classical, deterministic, poly-size circuits \(\{V_x : x \in \{0, 1\}^*\}\) and a polynomial \(p\), such that:

1. For all $x\in L_{\mathrm{yes}}$ there exists an $p(|x|)$-bit witness $w$ such that $V_x(w) = 1$;

2. For all $x\in L_{\mathrm{no}}$ and for all $p(|x|)$-bit witness $w$, $V_x(w) = 0$.

In the definition of NP, every candidate witness $w$ is a
classical bit string. For $x \in L_{\mathrm{no}}$, the
verifier must reject every classical witness candidate, and quantum states (quantum witnesses) are not considered within this framework.

\bigskip
\noindent	
\textbf{Quantum interactive proof systems for NP.}
A quantum interactive proof system for an NP relation $R$ consists of a pair of interactive machines $(P, V)$, where $P$ is the prover machine and $V$ is the verifier machine. In the multi-prover setting, the prover machine contains multiple spatially separated provers, and likewise the verifier machine may include multiple verifiers. The prover machine receives a classical input pair $(x, w)$, while the verifier receives only the input $x$.  In addition to the quantum soundness discussed in the main text, we recall below the formal definitions of completeness and quantum zero-knowledge for quantum proof systems for NP, following~\cite{Unruh2012quantum,watrous2006zeroknowledge}.

\begin{definition}[Completeness]
We call $(P, V)$ complete if there is a negligible function $\mu$ such that for all valid $(x, w)\in R$, we have that $\Pr[\left \langle P(x, w), V(x)\right \rangle = 1] \geq 1 - \mu(|x|)$.
When $\mu(|x|)=0$, we call it perfect completeness.
\end{definition}

\begin{definition}[Quantum zero-knowledge]
A proof system $(P,V)$ is quantum zero-knowledge if and only if for all polynomial-time verifiers $V^*$ there is a polynomial-time machine $S$ (the simulator) such that for all auxiliary quantum inputs $\rho$, and all $(x, w) \in R$, we have that the quantum state of malicious $V^*$ after an interaction $\left\langle P(x, w), V^*(x, \rho)\right\rangle$ is indistinguishable from the output of $S(x, \rho)$ of the simulator. If they are perfectly indistinguishable, we call it quantum perfect zero-knowledge.
\end{definition}

Note that although during the execution of the protocol the quantum-correlated provers may
share quantum states, employ arbitrary quantum strategies, and exchange quantum messages, the witness candidate $w$ for an NP relation remains classical by definition~\cite{regev2006quantum}. In contrast, in quantum proof systems for QMA the witness itself may be a quantum state, which leads to a fundamentally different setting and requires different techniques; such quantum witnesses are outside the scope of this work~\cite{Broadbent2016zkpQMA}.

\bigskip
\noindent	
\textbf{CHSH$_Q(2)$ game.} The $\text{CHSH}_Q(2)$ game generalizes the standard game $\text{CHSH}_2(2)$ to inputs from larger finite fields. Specifically, Alice and Bob receive uniformly random inputs \( x \in \mathbb{F}_Q \) and \( y \in \mathbb{F}_2 \), respectively, and output \( a, b \in \mathbb{F}_Q \). They win if \( a + b = x \cdot y \), where all arithmetic is in the finite field \( \mathbb{F}_Q \). 
  This game is projective with a uniform input distribution. While the exact classical and quantum values of \( \text{CHSH}_Q(2) \) remain unknown, constructing its coupled game enables one to derive an upper bound on the quantum value~\cite{chailloux2017relativistic,shi2024relativistic,weng2025tight}. In this case, the winning probability of the CHSH-type coupled game is easy to analyze due to the algebraic structure of CHSH games. (See the Supplementary Note 2.3 of the Supplementary Information for details). Therefore, we can obtain 
\begin{equation}
   \omega^*(\text{CHSH}_Q(2))\leq \frac{1}{2}+\frac{1}{\sqrt{2Q}}, 
\end{equation}
and this result coincides with the bound derived by ref.~\cite{sikora2014strong}. Although this bound is not tight, it is sufficient for certain cryptographic applications such as in our work to determine how many bits we should use for relativistic bit commitments.

\bigskip
\noindent	
\textbf{Proof of quantum soundness.}
Here we provide a detailed proof of Theorem~\ref{quantum_soundness}. As stated in Definition~\ref{quantum_soundness}, the goal of the proof is to show that when the input instance $x$ corresponds to a graph that is not three-colorable ($x\in L_{\mathrm{no}}$), any classical witness (color assignment) given to malicious quantum provers cannot pass the verifier's test with non-negligible probability even when quantum provers may share entanglement and perform arbitrary quantum operations during the commit and reveal phase.

Note that the soundness proof is carried out relative to arbitrary classical witnesses $w$, and does not attempt to derive or enforce a global labeling corresponding to the adversary's strategy. The protocol does not enforce that a cheating strategy behaves consistently with any global labeling across all edges, since each round checks just one edge and the commitments are not globally constrained. Therefore, one might concern that a general (possibly entangled) adversary may adopt contextual strategies whose behavior cannot be naturally described by any fixed classical color assignment (such as quantum witnesses). However, this does not weaken our result: our protocol only considers ZKP for NP relations, where the witness is inherently classical by definition. The witness $w$ is therefore used only as an analytic reference in the standard NP-soundness sense, rather than as a description of the general adversary's internal strategy. In addition, extending the framework to quantum witnesses would correspond to settings beyond NP, such as QMA-type problems, which are outside the scope of the present work.

We consider a NP oundness proof setting in which, at the beginning of each round, the provers are given an arbitrary classical witness $w=\{y_i\}_
{i\in V}$ (a color assignment). Because the graph is not three-colorable, every possible classical witness induces a partition of the edge set into two disjoint subsets $E = E_{\mathrm{c}} \cup E_{\mathrm{inc}}$ according to the structure of the graph, where $E_{\mathrm{c}}$ consists of edges whose vertices receive different colors, and $E_{\mathrm{inc}}$ consists of incorrectly colored edges whose endpoints receive the same color. Let $t = |E_{\mathrm{inc}}|$. Since the graph is not three-colorable, every candidate witness must satisfy $1 \le t \le |E|$.

(a) Case $\{i,j\}\in E_{\mathrm{c}}$.
When the challenged edge $\{i,j\}$ belongs to $E_{\mathrm{c}}$ (i.e., $y_i \neq y_j$), $P1$ and $P2$ can always win the round by simply following the protocol honestly. This occurs with probability $1 - \frac{t}{|E|}$. For such a properly colored edge, the provers can reveal the committed values in the reveal phase using the original encrypted keys $\{b_i, b_j\}$ corresponding to $\{y_i, y_j\}$, and always
satisfy both the consistency and proof checks.

(b) Case $\{i,j\}\in E_{\mathrm{inc}}$.
When the challenged edge $\{i,j\}$ lies in $E_{\mathrm{inc}}$ (i.e., $y_i = y_j$), which occurs with probability $\frac{t}{|E|}$, the provers must alter the committed colors in order to cheat. Without loss of generality, we assume $y_i = y_j = 2$, and the provers can fix the committed value of vertex $i$ to $2$ and try to reveal $y_j \in \{0,1\}$ in an attempt to pass the proof check in the check phase. However, by doing so, the quantum provers cannot always successfully pass the consistency check where $x_j\cdot y_j=a_j+b_j$.

This is due to the no-signaling principle: P1 does not know the challenge
selected by V2, and P2 does not know the input query $x_j$ sent by V1. The situation is equivalent to a nonlocal $\mathrm{CHSH}_Q(2)$ game: P1 receives $x_j\in\mathbb{F}_Q$ and outputs $a_j\in\mathbb{F}_Q$, while P2 receives $y_j\in\{0,1\}$ and outputs $b_j\in\mathbb{F}_Q$. The provers can successfully cheat (win the nonlocal game) if and only if $x_j\cdot y_j=a_j+ b_j$. The maximum cheating probability for quantum correlated provers who can share quantum states and adopt quantum strategies equals the quantum value of the nonlocal $\mathrm{CHSH}_Q(2)$ game, denoted by $\omega^*(\mathrm{CHSH}_Q(2))$, for which $\omega^*(\mathrm{CHSH}_Q(2)) \le \frac{1}{2} + \frac{1}{\sqrt{2Q}}$~\cite{sikora2014strong,shi2024relativistic,weng2025tight}.

\medskip
Therefore, combining the two cases, we obtain the single-round soundness error 
\begin{equation}
\begin{aligned}
\delta &= \Pr[\text{P1 and P2 pass the check phase} \mid \{i, j\} \in E_{\text{c}}]\\
&\quad+\Pr[\text{P1 and P2 pass the check phase} \mid \{i, j\}\in E_{\text{inc}}] \\
&\leq\max_{1\leq t\leq |E|}\left(1-\frac{t}{|E|}+\frac{t}{|E|}\cdot \omega^*\left(\text{CHSH}_Q(2)\right)\right)\\
&= 1 - \frac{1}{|E|} + \frac{1}{|E|} \cdot \omega^*\left(\mathrm{CHSH}_Q(2)\right) \\
&\leq 1 - \frac{1}{2|E|} + \frac{1}{\sqrt{2Q} \cdot |E|}.
\end{aligned}
\end{equation}

Choosing $Q = |E|^2$, i.e., $\log Q=2\log |E|$, we have $\frac{1}{\sqrt{2Q}|E|} \ll \frac{1}{2|E|}$, and the term can be neglected. For $m = 2k|E|$ rounds, the overall soundness error satisfies
\begin{equation}
    \begin{aligned}
        \delta_s &\leq \left(1 - \frac{1}{2|E|} + \frac{1}{\sqrt{2Q}|E|}\right)^m \\
        &=\sum_{s=0}^m\binom{m}{s}\left(1-\frac{1}{2|E|}\right)^{s}\cdot \left(\frac{1}{\sqrt{2Q}|E|}\right)^{m-s}\\
        &\approx \left(1 - \frac{1}{2|E|} \right)^{m} \\
        &= \left(1 - \frac{1}{2|E|} \right)^{2k|E|} \\
        &\approx e^{-k},
    \end{aligned}
\end{equation}
assuming $|E|$ is sufficiently large.

\bigskip
\noindent	
\textbf{{Perfect completeness.}}
If the graph $\mathbb{G}$ is three-colorable and the provers honestly follow the protocol, then regardless of the edge $C$$=$$\{i, j\}$ chosen by V2, in our relativistic GMW ZKP, the verifiers will definitely observe $y_i$$\neq$$y_j$ and accept. It can be formally written as
\begin{theorem}[Perfect completeness of asymmetric RZKP]\label{thm:perfect_completeness}
    If the graph $\mathbb{G}$ is three-colorable ($x\in L_{yes}$), and P1 and P2 honestly follow the protocol, then $\Pr[\text{\rm V accepts $x$}]=1$.
\end{theorem}
The proof of Theorem~\ref{thm:perfect_completeness} can be found in Supplementary Note 5.1 of the Supplementary Information.

\bigskip
\noindent
\textbf{Quantum perfect zero-knowledge.} 
To formally establish the zero-knowledge property, a simulator must be constructed that operates without any prior knowledge and generates a view for the verifiers that is indistinguishable from the one produced during actual interactions. Proving zero-knowledge against quantum verifiers with auxiliary quantum states poses significant challenges, as it typically requires the simulator to perform complex quantum rewinding~\cite{watrous2006zeroknowledge}.

Interestingly, quantum rewinding is unnecessary in RZKPs. The simple mathematical structure of relativistic bit commitment enables the simulator, operating in the ideal world without relativistic constraints, to reveal any colors for the challenged edge as desired, all within polynomial time. This virtual process is fundamentally impossible for real provers due to the binding property of relativistic bit commitment, which prevents them from altering their committed values. This asymmetry arises from the relativistic constraints imposed on the provers~\cite{chailloux2017relativistic,crepeau2023zeroknowledge}. Moreover, in both real interactions and simulations, the relativistic bit commitment ensures perfect hiding of the colors, preventing any information leakage to the verifiers. The simulator can therefore efficiently simulate each round of the protocol sequentially, from the first to the last, generating a view that is perfectly identical to that of the real interaction, without requiring any knowledge of the actual three-colorings.

\begin{theorem}[Quantum perfect zero-knowledge of asymmetric RZKP]\label{thm:QZK}
Our asymmetric RZKP protocol is quantum perfect zero-knowledge, which means that the simulator successfully simulates a view that is perfectly indistinguishable from that of the real interaction without the valid witness $Y_{\pi}$. This can be formally expressed as:  
\begin{equation}
\begin{aligned}
&\forall x \in L_{\text{yes}}, \forall \pi \in \Pi, \forall X \in \mathbb{F}_{Q}^{\otimes |V|}, \forall C \in E, \forall \text{poly-qubit} \, \rho, \\
&\sigma^f_{\text{real}}[P(x, Y_\pi) \leftrightarrow V^*(x, X, C, \rho)] = \sigma^f_{\text{sim}}(x, X, C, \rho).
\end{aligned}
\end{equation}   
\end{theorem}
The proof of Theorem~\ref{thm:QZK} can be found in Supplementary Note 5.2 of the Supplementary Information. 

\bigskip
\noindent
\textbf{Scalability of RZKPs.}  
Our proof-of-principle experimental implementation demonstrates the feasibility of asymmetric RZKPs. Nevertheless, several key challenges remain for scaling RZKPs to practical real-life fiber networks. Below we outline the main limitations and potential solutions for large-scale deployment. Points 1 and 2 are common to all RZKP protocols, while Point 3 is specific to our asymmetric construction.

1. Distance between prover and verifier: In our protocol, we assume that P1 and V1 (and likewise P2 and V2) are co-located during execution, and that V1 and V2 are aware of the true locations of P1 and P2. However, in large-scale fiber-optical networks, this assumption may not hold: for example, P1 could attempt to mislead V1 about their location while actually being co-located with P2. One possible quantum-secure solution is to integrate quantum position verification (QPV) protocols to authenticate P1's location before the ZKP execution. Although QPV is theoretically feasible, practical deployment still faces significant challenges, such as reliable three-dimensional position verification, precise compensation of fiber delays, and constraints imposed by network topology. Therefore, integrating QPV with RZKPs remains an open problem for future research.

2. Number of prover and verifier agents: In RZKPs, the system can be understood as involving only two actual users: a prover and a verifier.  P1 and P2 (resp.~V1 and V2) are simply trusted agents of the prover (resp.~verifier). At least two prover agents are necessary in relativistic cryptographic protocols, since the no-signaling principle, which ensures security, would not hold with only one prover agent. In principle, the verifier side could be reduced to a single agent (e.g., letting V1 communicate separately with P1 and P2 while accounting for time delays). In practice, however, this introduces significant challenges, such as the need for precise spatial separation of P1 and P2 and strict control of time delays in the fiber. For these reasons, the two-prover/two-verifier configuration may represent the minimal and most practical setting for ensuring the security of relativistic cryptographic protocols.

3. Computational load in our asymmetric protocol: As discussed in the main text, our asymmetric protocol achieves significant reductions in round complexity and overall randomness cost. Nevertheless, the P1-V1 pair bears the majority of the computational and randomness-generation burden, as they are responsible for encrypting the colors of all vertices. This may present practical challenges for implementation when the graph contains a large number of vertices and edges. We hope that future work can design an RZKP protocol that simultaneously achieves low round complexity and low per-round randomness cost.

\bigskip
\noindent
\textbf{Note added to the discussion on scalability.} 
During the revision of this work, we became aware of a new study by Yao \textit{et al.}~\cite{ma2025quantum}, which explores the practical implementation of symmetric RZKPs in both the two-prover setting (classical soundness) and the three-prover setting (quantum soundness) from an engineering perspective. In addition, it resolves the two-prover quantum soundness property for symmetric RZKP protocols, which was previously left as an open problem in earlier works~\cite{crepeau2019practical,alikhani2021experimental}. For readers seeking a more comprehensive discussion on the practical implementation, scalability, and potential real-world applications of RZKPs, we recommend referring to their work.

\bigskip
\noindent
\textbf{ZKP and ZKPoK.}
Zero-knowledge proofs (ZKPs) and zero-knowledge proofs of knowledge (ZKPoKs) are closely related but conceptually distinct notions in cryptography. A ZKP allows a prover to convince a verifier that a given statement is true without revealing any additional information. In contrast, a ZKPoK further guarantees that the prover actually possesses a valid witness (e.g., a secret key or a valid solution) for the statement being proven. This distinction is particularly crucial in applications such as identity verification, where it is not sufficient for a prover to demonstrate the existence of a valid credential: the verifier must be assured that the prover actually knows it. Briefly, a standard ZKP proves the existence of a witness, while ZKPoK proves that the prover actually knows the witness.

In our present work, we focus on establishing quantum soundness for a RZKP protocol for graph three-coloring. While our protocol preserves quantum zero-knowledge and offers soundness against quantum correlated adversaries, it does not yet establish knowledge soundness as required by ZKPoK. As a result, applying our current protocol and previous RZKP protocols to identity verification would be premature. Nevertheless, the structure of our asymmetric RZKP may provide a useful foundation for developing relativistic ZKPoK protocols. We leave this as an open direction for further investigation.

\bigskip
\noindent	
\begin{center}
    \textbf{\large Acknowledgements}
\end{center}

We appreciate the support from Nanjing University-China Mobile Communications Group Co., Ltd. Joint Institute. We thank Marco Tomamichel and Minglong Qin for their valuable discussions on the security analysis. We also thank Zhejiang Saisi Electronic Technology Co., Ltd. for providing high-precision GPS time synchronization servers. C.-X.W. appreciates the hospitality of the Centre for Quantum Technologies at the National University of Singapore.

We gratefully acknowledge the creators and providers of all free graphic resources for commercial use in this work. The verifier and prover in Fig.~\ref{fig1} and Supplementary Fig.~3 are the \href{https://www.flaticon.com/free-sticker/service_8662591}{service} and \href{https://www.flaticon.com/free-sticker/social-media_8662598}{social-media} stickers created by Kerismaker - Flaticon. The satellite in Fig.~\ref{fig3}a is the \href{https://www.flaticon.com/free-sticker/satellite_12142417}{satellite} sticker created by Design Circle - Flaticon. The antenna in Fig.~\ref{fig3}a is the \href{https://www.flaticon.com/free-sticker/signal_9821271}{signal} sticker created by vectorsmarket15 - Flaticon. The laptop in Fig.~\ref{fig3}a is the \href{https://www.flaticon.com/free-sticker/laptop_11483681}{laptop} sticker created by Stickers - Flaticon. The GPS time server and FPGA shown in Fig.~\ref{fig3} are real photographs taken during the actual experiment. The aerial photograph of the Xianlin Campus of Nanjing University was captured using a drone. The quantum entanglement in Supplementary Fig.~2 is the \href{https://www.flaticon.com/free-icon/atom_3066074}{atom} icon created by FauzIDEA - Flaticon. The simulator in Supplementary Fig.~3 is the \href{https://www.flaticon.com/free-sticker/robot_12239721}{robot} sticker created by Stickers - Flaticon.

\bigskip
\noindent	
\begin{center}
    \textbf{\large Funding}
\end{center}
This work was supported by the National Natural Science Foundation of China (Nos. 12522419, U25D8016, 12274223 and 12475020), the Program for Innovative Talents, Entrepreneurs in Jiangsu (No. JSSCRC2021484), the Fundamental Research Funds for the Central Universities and the Research Funds of Renmin University of China (No. 24XNKJ14), the National Key Research and development Program of China (No. 2023YFC2205802), the Innovation Program for Quantum Science and Technology (No. 2021ZD0301701), and the China Scholarship Council (No. 202406190220).

\bibliographystyle{naturemag}
%\bibliographystyle{apsrev}
%%%%%%%%%%%%%%%%%%%%%%%%%%%%%%%%%%%%%%%
% choose a .bib file
%\bibliography{reftext}

\end{document}

\title{Supplementary Information: Experimental asymmetric relativistic zero-knowledge proofs with unconditional security}

\author{Chen-Xun Weng}
\affiliation{National Laboratory of Solid State Microstructures and School of Physics, Collaborative Innovation Center of Advanced Microstructures, Nanjing University, Nanjing 210093, China}
\affiliation{School of Physics and Beijing Key Laboratory of Opto-electronic Functional Materials and Micro-nano Devices, Key Laboratory of Quantum State Construction and Manipulation (Ministry  of  Education), Renmin University of China, Beijing 100872, China}
\affiliation{Centre for Quantum Technologies, National University of Singapore, Singapore 117543, Singapore}
\author{Ming-Yang Li}
\affiliation{National Laboratory of Solid State Microstructures and School of Physics, Collaborative Innovation Center of Advanced Microstructures, Nanjing University, Nanjing 210093, China}
\affiliation{School of Physics and Beijing Key Laboratory of Opto-electronic Functional Materials and Micro-nano Devices, Key Laboratory of Quantum State Construction and Manipulation (Ministry  of  Education), Renmin University of China, Beijing 100872, China}
\author{Nai-Rui Xu}
\affiliation{MatricTime Digital Technology Co. Ltd., Nanjing 211899, China}
\author{Yanglin Hu}
\author{Ian George}
\author{Jiawei Wu}
\affiliation{Centre for Quantum Technologies, National University of Singapore, Singapore 117543, Singapore}
\author{Shengjun Wu}\email{sjwu@nju.edu.cn}
\affiliation{National Laboratory of Solid State Microstructures and School of Physics, Collaborative Innovation Center of Advanced Microstructures, Nanjing University, Nanjing 210093, China}
\author{Hua-Lei Yin}\email{hlyin@ruc.edu.cn}
\affiliation{School of Physics and Beijing Key Laboratory of Opto-electronic Functional Materials and Micro-nano Devices, Key Laboratory of Quantum State Construction and Manipulation (Ministry  of  Education), Renmin University of China, Beijing 100872, China}
\affiliation{National Laboratory of Solid State Microstructures and School of Physics, Collaborative Innovation Center of Advanced Microstructures, Nanjing University, Nanjing 210093, China}
\author{Zeng-Bing Chen}\email{zbchen@nju.edu.cn}
\affiliation{National Laboratory of Solid State Microstructures and School of Physics, Collaborative Innovation Center of Advanced Microstructures, Nanjing University, Nanjing 210093, China}
\affiliation{MatricTime Digital Technology Co. Ltd., Nanjing 211899, China}

\date{\today}
\maketitle

\tableofcontents
\newpage

\section{Supplementary Note 1: Mathematical preliminaries}
\subsection{1.1 Finite fields}
A finite field (or Galois field) is denoted as $\mathbb{F}_Q$ or $\text{GF}(Q)$, where $Q = q^k$, with $q$ being a prime number (the characteristic of the field), and $k$ being a positive integer. The finite field contains $Q$ elements. Finite fields have the following key properties:
\begin{itemize}
    \item Closure: For any two elements $a, b\in \mathbb{F}_Q$, the results of addition, subtraction, multiplication, and division (except division by zero) remain within $\mathbb{F}_Q$.

    \item Associativity: For any elements $a, b, c \in \mathbb{F}_Q$, the following hold: $(a + b) + c = a + (b + c)$ and $(a \cdot b) \cdot c = a \cdot (b \cdot c)$.

    \item  Commutativity: For any elements $a, b \in \mathbb{F}_Q$, the following hold: $a + b = b + a$, $a \cdot b = b \cdot a$.

    \item Distributivity: For any elements $a, b, c \in \mathbb{F}_Q$, the following holds: $a \cdot (b + c) = a \cdot b + a \cdot c$.

    \item Additive Identity: There exists an element  $0 \in \mathbb{F}_Q$ such that for any $ a \in \mathbb{F}_Q $, we have: $a + 0 = a$.

    \item Multiplicative Identity: There exists an element $ 1 \in \mathbb{F}_Q$ such that for any $a \in \mathbb{F}_Q$, we have: $a \cdot 1 = a$.

    \item Additive Inverses: For each $a \in \mathbb{F}_Q$, there exists an element $-a \in \mathbb{F}_Q$ such that: $a + (-a) = 0$.

    \item Multiplicative Inverses: For each nonzero element $a \in \mathbb{F}_Q$, there exists an element $a^{-1} \in \mathbb{F}_Q$ such that: $a \cdot a^{-1} = 1$.

    \item Prime field $\mathbb{F}_q$: When $Q = q$ (i.e., $k = 1$), the field consists of the integers $\{0, 1, 2, \dots, q-1\}$, and addition and multiplication are performed modulo $q$.

    \item Extension Field $\mathbb{F}_{Q}$: When $k > 1$, the field consists of polynomials over $\mathbb{F}_q$ with degree less than $k$, and arithmetic is performed modulo an irreducible polynomial of degree $k$. It is constructed as the quotient ring: 
    \begin{equation}
        \mathbb{F}_{q^k} = \mathbb{F}_q[x] / (f(x)),
    \end{equation}
    where $f(x)$ is an irreducible polynomial of degree $k$ over $\mathbb{F}_q$. The elements of $\mathbb{F}_{q^k}$ can be represented as polynomials of degree less than $k$ with coefficients in $\mathbb{F}_q$. For example, in $\mathbb{F}_{2^3}$, we can choose $f(x) = x^3 + x + 1$. Note that the irreducible polynomial $f(x)$ used to construct the extension field $\mathbb{F}_{q^k}$ is not unique, as there are multiple irreducible polynomials of degree $k$ over $\mathbb{F}_q$. Each of these polynomials can serve as the modulus for defining $\mathbb{F}_{q^k}$, resulting in fields that are all isomorphic, meaning they share the same algebraic structure despite being represented differently. The number of such irreducible polynomials is given by $N_q(k) = \frac{1}{k} \sum_{d \mid k} \mu(d) q^{k/d}$, where $d\mid k$ means the sum is taken over all divisors $d$ of $k$ and $\mu(d)$ is the M$\rm \ddot{o}$bius function. For example, in $\mathbb{F}_{2^3}$, both $f(x) = x^3 + x + 1$ and $f(x) = x^3 + x^2 + 1$ are irreducible polynomials, and either can be used to define the field.   
    \begin{itemize}
        \item[0.] Representation of Elements: elements of $ \mathbb{F}_{q^k} $ are of the form: $a(x) = a_0 + a_1x + a_2x^2 + \cdots + a_{k-1}x^{k-1}$, where $a_i \in \mathbb{F}_q$.
        \item[1.] Addition and Subtraction: addition and subtraction are all performed component-wise: $a(x)\pm b(x) = \sum_{i=0}^{k-1} (a_i \pm b_i)x^i$, where addition and subtraction of coefficients are done in $\mathbb{F}_q$.
        \item[2.] Multiplication: multiplication involves multiplying two polynomials in $\mathbb{F}_q[x]$ and reducing modulo $f(x)$: $c(x) = a(x) \cdot b(x) \mod f(x)$.
        \item[3.] Inverse (Division): To compute the inverse of $b(x)$, use the extended Euclidean algorithm to find $b(x)^{-1}$ such that: $b(x) \cdot b(x)^{-1} \equiv 1 \mod f(x)$.        
    \end{itemize}
\end{itemize}

\subsection{1.2 Graph three-coloring problem}
Three-colorability of a graph $\mathbb{G}(V,E)$, where $V$ and $E$ are the sets of vertices and edges respectively, is formally defined as follows:

\begin{definition}
    A graph $\mathbb{G}(V,E)$ is three-colorable if its vertices can be colored with only three colors, such that no two vertices of the same color are connected by an edge, i.e., $\forall \{u,v\}\in E$, $y_u\neq y_v$.
\end{definition}

\begin{figure}[h]
    \centering
    \includegraphics[width=0.8\linewidth]{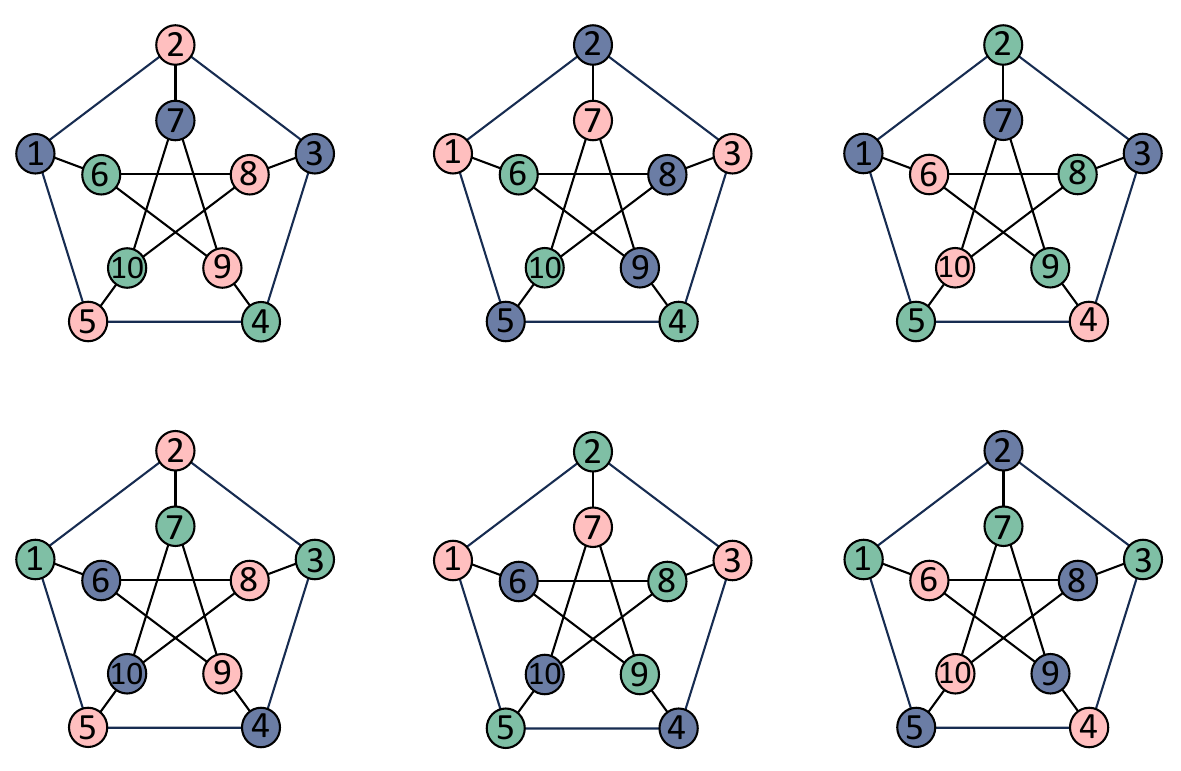}
    \caption{\textbf{A three-colorable graph with possible color permutations.} Three-colorability requires that each vertex in a graph is assigned one of three colors such that no two adjacent vertices share the same color. This  example graph, containing ten vertices, satisfies this condition, as illustrated by its three-coloring using blue, pink, and green. All adjacent vertices such as 1 and 6, or 8 and 10, are assigned different colors. Any permutation of a valid three-coloring is also valid. The permutations of blue, pink, and green coloring provide equivalent correct solutions. Thus, there are at least $|\Pi|=6$ such cyclic permutations.}
    \label{fig:3-coloring}
\end{figure}

Supplementary Figure~\ref{fig:3-coloring} illustrates a three-colorable graph and its possible color permutations. The three-coloring problem, an NP-complete problem, holds significant implications for the design of ZKPs~\cite{alikhani2021experimental,li2023device}. This problem asks whether a given graph is three-colorable, i.e., whether the vertices of the graph can be colored using only three colors such that no two adjacent vertices share the same color. All NP problems can be converted to the three-coloring problem with a polynomial-time algorithm~\cite{arora2009computational}, ensuring that a ZKP protocol for three-coloring readily extends to all problems within the NP complexity class. This inherent computational intractability motivates the development of efficient and secure ZKP protocols for this problem. As shown in Supplementary Fig.~\ref{fig:3-coloring}, any permutation of a valid three-coloring remains a valid solution.  Therefore, the number of possible three-colorings is at least $|\Pi| = 6$ for any three-colorable graph, corresponding to the six cyclic permutations of three colors.

\section{Supplementary Note 2: Non-local game}
Non-local entanglement games are cooperative quantum games involving multiple spatially separated players who share entangled quantum states. The no-signaling principle prevents direct communication between players. Each player receives a random input and generates an output based on a local measurement of their subsystem of the entangled state. The characteristic of such games is the players' ability to exhibit correlations that surpass the capabilities of any classical strategy. This section firstly introduces the no-signaling principle. A formal definition of a non-local game is then provided, along with a method for constructing coupled games to determine upper bounds on the quantum winning probability. This method is subsequently applied to derive upper bounds for the quantum winning probability of the CHSH$_Q(P)$ game. %and and its $n$-fold parallel repetition. 

\subsection{2.1 No-signaling principle}
The no-signaling principle is a fundamental concept of modern physics, which states that during the measurement of an entangled quantum state, it is impossible for one observer to transmit information to another observer, regardless of their spatial separation. This conclusion preserves the principle of causality in quantum mechanics and ensures that information transfer does not violate special relativity by exceeding the speed of light~\cite{Peres2004Quantum}. In relativistic cryptography, this principle is essential to preclude any form of coordinated cheaters, regardless of the presence of shared entanglement~\cite{Kalai2023quantum}.

\subsection{2.2 Non-local Game $G$ and its coupled game $G_{\text{coup}}$}
A non-local game $G$ is defined by the tuple $(I_A, I_B, O_A, O_B, V, p)$, where $I_A$ and $I_B$ are the input sets for Alice and Bob respectively; $O_A$ and $O_B$ are their corresponding output sets; $V: I_A \times I_B \times O_A \times O_B \to \{0, 1\}$ is the verification function, with $V(x, y, a, b) = 1$ signifying a win and 0 a loss for inputs $x, y$ and outputs $a, b$; and $p: I_A \times I_B \to [0, 1]$ is the input distribution, satisfying $\sum_{(x,y) \in I_A \times I_B} p(x, y) = 1$.

\begin{definition}[Uniform distribution]
    A non-local game $G = (I_A, I_B, O_A, O_B, V, p)$ is defined to be on the uniform distribution if $p(x,y) = \frac{1}{|I_A||I_B|}$ for all $(x,y) \in I_A \times I_B$.
\end{definition}

\begin{definition}[Projective]
     A game $G=(I_A, I_B, O_A, O_B, V, p)$ is projective if for any $(x, y, a)\in I_A\times I_B\times O_A$, there is a unique $b\in O_B$ such that $V(x,y,a,b) = 1$.
\end{definition}

\begin{definition}[Winning probability] 
    For a game $G=(I_A, I_B, O_A, O_B, V, p)$, we denote its classical winning probability value as $\omega(G)$ if Alice and Bob are classical correlated and only adopt classical strategies, and denote its quantum winning probability value as $\omega^*(G)$ if Alice and Bob are quantum correlated and can adopt quantum strategies. 
\end{definition}

Determining the exact quantum value, or even a tight upper bound, on the winning probability of a non-local game is generally challenging. However, a method for constructing a coupled game to obtain an upper bound on the quantum value for a given non-local game $G$ has been proposed in refs.~\cite{chailloux2017relativistic,Unruh2012quantum,shi2024relativistic,weng2025tight}, which gives an intriguing general method to find the upper bound on quantum value by constructing the coupled game of $G$. The following context details the construction of such coupled game $G_{\text{coup}}$ for any non-local game $G$.

\begin{definition}[The construction of $G_{\text{coup}}$]
    For any game $G=(I_A, I_B, O_A, O_B, V, p)$  defined on a uniform input distribution, its coupled game $G_{\text{coup}}$ is constructed as follows:
    \begin{itemize}
        \item[1.] Alice receives a uniformly random input $x\in I_A$, while Bob receives a uniformly random pair of distinct inputs $y, y^{\prime}\in I_B$ such that $y\neq y^{\prime}$.
        \item[2.] Alice outputs $a\in O_A$, and Bob outputs $b, b^{\prime}\in O_B$.
        \item[3.] Alice and Bob win this coupled game if and only if $V(x,y,a,b)=V(x,y^{\prime},a,b^{\prime})=1$.The quantum winning probability of this coupled game is denoted by $\omega^*(G_{\text{coup}})$.
    \end{itemize}
 
\end{definition}
The scheme of a non-local game $G$ and its coupled game $G_{\rm{coup}}$ are illustrated in Supplementary Fig.~\ref{fig:non-local game}. The tightest relationship between the quantum winning probabilities of $G$ and $G_{\text{coup}}$ can be established using the consecutive measurement theorem~\cite{weng2025tight}.

\begin{figure}
    \centering
    \includegraphics[width=0.8\linewidth]{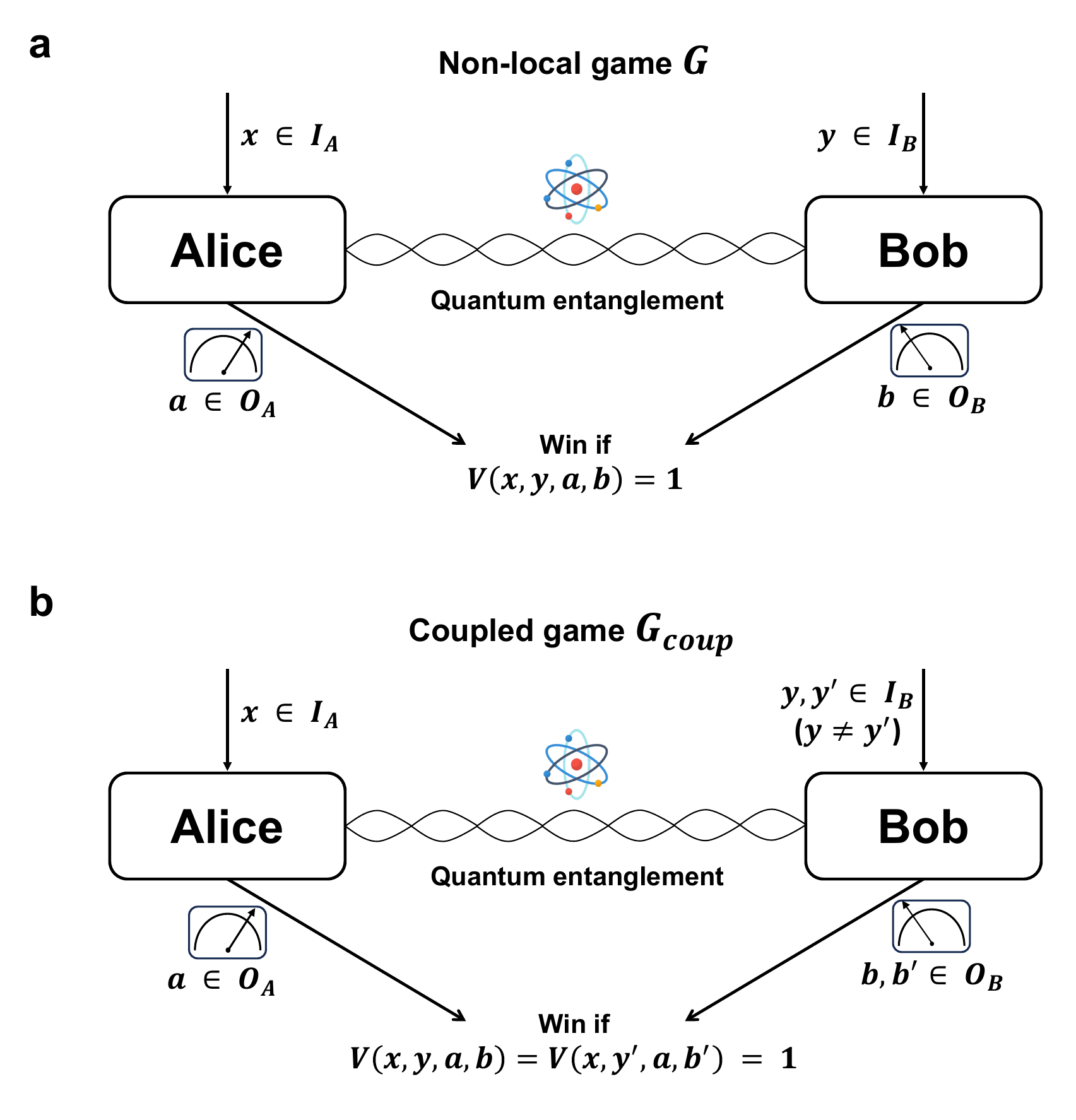}
    \caption{\textbf{Illustration of a non-local game $G$ and its coupled game $G_{\rm{coup}}$.} \textbf{(A) Non-local game $G$.} The physical systems of Alice and Bob share quantum entanglement. Alice gets an input $x \in I_A$ and Bob gets an input $y \in I_B$. Based on the received input, Alice and Bob choose a measurement on their systems and generate the output $a \in O_A$ and $b \in O_B$ respectively. They win the game if the valuation function produces 1, i.e., $V(x,y,a,b)=1$. \textbf{(B) Coupled game $G_{\rm{coup}}$.} In the coupled game, Alice maintains the same input and output condition, but Bob gets two distinct inputs $y,y^{\prime} \in I_B~(y \neq y^{\prime})$. Bob also generates two outputs $b,b^{\prime} \in O_B$, subsequent to measurement on his physical system. Alice and Bob win if $V(x,y,a,b)=V(x,y^{\prime},a,b^{\prime})=1$.}
    \label{fig:non-local game}
\end{figure}

\begin{theorem} [The relationship between $G$ and $G_{\text{coup}}$]\label{G_coup_relationship}
For any projective non-local game $G$ whose inputs are on the uniform distribution, we have 
\begin{equation}
    \omega^*(G_{\text{coup}})\ge \left(\frac{|I_B|}{|I_B|-1}\right)^2\omega^{*}(G)\cdot\left(\omega^*(G)-\frac{1}{|I_B|}\right)^2,
\end{equation}
where $|I_B|$ is the dimension of Bob's input size.
\end{theorem}

While the exact quantum winning probability, or even a tight upper bound, is often difficult to determine for certain non-local games, the upper bound for their corresponding coupled games is readily obtainable via the no-signaling principle~\cite{chailloux2017relativistic}. This allows us to establish upper bounds for the original non-local games through Theorem~\ref{G_coup_relationship}. The following examples demonstrate this approach for the non-local CHSH$_Q(2)$ game. Although the bound may not be the tightest possible, they provide sufficient security for various cryptographic applications, including relativistic bit commitment protocols.

Here we only show the main results of ref.~\cite{weng2025tight}. Please see ref.~\cite{weng2025tight} for the detailed mathematical proof of Theorem~\ref{G_coup_relationship}.

\subsection{2.3 CHSH$_Q(2)$ game}\label{CHSH}

We have defined two key parameters for any non-local game $G$: the classical winning probability $\omega(G)$ and the quantum winning probability $\omega^*(G)$.  To illustrate, consider the standard binary CHSH game, CHSH$_2(2)$. In this game, two players, Alice and Bob, each receive a uniformly random bit ($x, y \in \mathbb{F}_2 = \{0, 1\}$) as input.  They each output a bit ($a, b \in \mathbb{F}_2 = \{0, 1\}$). They win if $a + b = x \cdot y $. For CHSH$_2(2)$, the classical winning probability is $\omega(\text{CHSH}_2(2)) = 0.75$, while the quantum winning probability is $\omega^*(\text{CHSH}_2(2)) = \frac{1}{2} + \frac{\sqrt{2}}{4} \approx 0.85$. The optimal quantum strategy achieving these bounds is known for both classical and quantum scenarios.

The CHSH$_Q(2)$ game generalizes the CHSH$_2(2)$ game to inputs and outputs to larger finite fields. Specifically, Alice and Bob receive uniformly random inputs $x \in \mathbb{F}_Q$ and $y \in \mathbb{F}_2$, respectively, and output $a \in \mathbb{F}_Q$ and $b \in \mathbb{F}_Q$.  They win if $a + b = x \cdot y$, where arithmetic is performed in the finite field $\mathbb{F}_Q$. This game is 1-projective ($S=1$) under a uniform input distribution; that is, for given $x$, $a$, and $y$, there exists only a unique $b$ satisfies $a + b = x\cdot y$.  While the optimal classical and quantum winning probabilities for CHSH$_Q(2)$ remain unknown, and determining even upper bounds is challenging, constructing its coupled game allows us to derive an upper bound on the quantum winning probability~\cite{shi2024relativistic,weng2025tight}. Although not necessarily tightest, this bound is sufficient for certain cryptographic applications.

\begin{theorem}[Upper bound of $\omega^{*}$(CHSH$_{Q}(2)$)]\label{CHSH_upper}
    For integer $Q\geq 2$, the upper bound on the quantum winning probability of the CHSH$_Q(2)$ game is given by:
\begin{equation}
    \omega^*(\text{CHSH}_Q(2)) \le \frac{1}{2} + \frac{1}{\sqrt{2Q}}.
\end{equation}
\end{theorem}

\textit{Proof.} We construct the coupled game CHSH$_Q(2)_\text{coup}$ as follows.  First, fix Alice's input/output pair $(x, a)$ and randomly select two distinct inputs $y, y^{\prime}$ for Bob. Bob then outputs $(b, b^{\prime})$ corresponding to these inputs.  A win in CHSH$_Q(2)_\text{coup}$ requires that both $V(x, y, a, b) = 1$ and $V(x, y^{\prime}, a, b^{\prime}) = 1$, implying:
\begin{equation}
    a+b=x\cdot y \And a+b^{\prime}=x\cdot y^{\prime}.
\end{equation}
This leads to $x = \frac{b - b^{\prime}}{y - y^{\prime}}$. Thus, to win CHSH$_Q(2)_\text{coup}$, Bob must correctly guess Alice's input $x$, an event with probability at most $\frac{1}{Q}$ due to the no-signaling principle. Then we have:
\begin{equation}
    \omega^*(\text{CHSH}_Q(2)_\text{coup}) \le \frac{1}{Q}.
\end{equation}
Applying Theorem \ref{G_coup_relationship} with $|I_B| = 2$, we obtain the upper bound on the quantum winning probability of the CHSH$_Q(2)$ game:
\begin{equation}
    4\omega^*(\text{CHSH}_Q(2))\left(\omega^*(\text{CHSH}_Q(2))-\frac{1}{2}\right)^2 \le \frac{1}{Q}
\end{equation}
Considering the fact that $\omega^*(\text{CHSH}_Q(2))\geq\frac{1}{2}$, i.e., $4\omega^*(\text{CHSH}_Q(2))\geq 2$, we obtain
\begin{equation}
   2\left( \omega^*(\text{CHSH}_Q(2))-\frac{1}{2}\right)^2 \le \frac{1}{Q}
\end{equation}
Therefore, we can prove that
\begin{equation}
    \omega^*(\text{CHSH}_Q(2)) \le \frac{1}{2} + \frac{1}{\sqrt{2Q}},
\end{equation}
and this result coincides with that of ref.~\cite{sikora2014strong}.

\bigskip 
\section{Supplementary Note 3: Relativistic bit commitment}

\subsection{3.1 Two properties: hiding and sum-binding}
We now demonstrate how the relativistic protocol satisfies the two fundamental properties of bit commitment: hiding (or concealing) and binding.  In this context,  provers P1 and P2 each possess a bit, $y$, which they wish to commit to a verifier.  The hiding property requires that the verifier cannot learn the value of $y$ prior to its revelation by the provers. Simultaneously, the binding property ensures that the provers cannot change the value of $y$ after the commitment phase. While unconditionally secure quantum bit commitment is known to be impossible \cite{lo1997isquantum,mayers1997unconditionally},  relativistic bit commitment leverages the principles of special relativity to circumvent this limitation.

\begin{definition}[$\mathbb{F}_Q$ relativistic two-prover bit commitment]
A relativistic commitment scheme is defined by the interactive protocol, involving the commit and open strategies denoted as (Com, Open), between two verifiers and two provers.

\begin{itemize}
    \item[0.] Preparation phase: P1 and P2, separated by a distance $D$,  jointly generate a uniformly random encoding key $b$. 
    \item[1.] Commit phase: V1 sends a uniformly random query $x \in \mathbb{F}_Q$ to P1.  P1 computes $a = x\cdot y- b$, where $y\in \mathbb{F}_P$ ($Q>P$) is the ``bit" to be committed. P1 sends $a$ to V1. 
    \item[2.] Reveal phase: P2 sends the key $b$ and the committed bit $y$ to V2. V2 verifies the commitment by checking if $a = x\cdot y -b$. The commitment is accepted if the equation holds; otherwise, it is rejected. 
\end{itemize}
\end{definition}

The commitment need not be restricted to a binary value. The committed value, $y$, may represent multiple values or a set of bits; for notational simplicity, we refer to this committed value as a "bit", regardless of the number of possible values it can take.

\subsection{3.2 Hiding}

The hiding (concealing) property ensures that the committed bit remains hidden from the verifier until the reveal phase. This implies that guessing the bit is computationally or physically infeasible. Formally, this is expressed as indistinguishability between the possible committed bit values. Relativistic bit commitment schemes could achieve perfect hiding, meaning the verifier gains no information about the committed bit before the reveal phase~\cite{shi2024relativistic}.

\begin{definition}[Perfect hiding (concealing)]
 Perfect hiding ensures that the commitment reveals no information about the committed value $y$ to the verifier before the reveal phase. Let $\mathcal{A}(y, b)$ denote the commitment generated for a value $y \in \{0, 1, \dots, P-1\}$ using randomness $b$, and let $\mathcal{A}(y)$ represent the distribution of commitments for all possible choices of $b$. The scheme is perfectly hiding if:
\begin{equation}
  \mathcal{A}(y) \overset{d}{=} \mathcal{A}(y'), \quad \forall y, y' \in \{0, 1, \dots, P-1\},  
\end{equation}
where $ \overset{d}{=} $ denotes perfect equality in distribution. 
\end{definition}

\subsection{3.3 Sum-binding and its limitations}

In the context of relativistic bit commitment, the sum-binding property is a measure of how well the protocol resists cheating by the provers. Specifically, it quantifies the provers' inability to alter the committed bit after the commit phase, even if the provers share quantum entanglement and use quantum strategies. The sum-binding property is defined as follows.

\begin{definition}[Sum-binding]
In a relativistic setting with quantum correlated provers sharing quantum entanglement, a commitment scheme is sum-binding if, for all possible malicious quantum strategies, (Com$^*$, Open$^*$), employed by the provers during the commitment and reveal phases, the following inequality holds:
\begin{equation}
    \forall \text{Com}^*, \sum_y \max_{\text{Open}^*}\Pr[\text{P1 and P2 successfully reveal \textit{y}} \, | \, (\text{Com}^*, \text{Open}^*)] \le 1+\varepsilon_b.
\end{equation}
where $\varepsilon_b$ represents the binding error and it is a positive negligibly small quantity.

\end{definition}

Relativistic bit commitment protocols are often characterized by the security notion of sum-binding, as defined in Ref.~\cite{kaniewski2013secure}. For a commitment to a value from a set $\mathcal{C}$, this property asserts that the sum of the probabilities, $p_c$, of a malicious party successfully opening the commitment to any value $c \in \mathcal{C}$ is bounded: $\sum_{c \in \mathcal{C}} p_c \leq 1 + \varepsilon_b$, for a small security parameter $\varepsilon_b$. This property constrains the total probability of a successful opening, but it imposes no non-trivial bound on the individual probabilities. An adversary is thus not precluded from possessing a non-negligible success probability for multiple distinct outcomes. This inherent ambiguity in the commitment is the reason that sum-binding is not a composable security definition.

The failure of composability is demonstrated by constructing a string commitment protocol from $n$ parallel instances of a sum-binding bit commitment. An adversary can employ a global strategy that introduces correlations across all $n$ instances. Consider an attack where with probability 1/2, the adversary can perfectly open each of the $n$ bits to either `0' or `1' independently, and with probability 1/2, the attempt fails entirely. From the perspective of any single bit commitment, the probability of successfully opening to `0' is $p_0 = 0.5$ and to `1' is $p_1 = 0.5$, satisfying the sum-binding condition $p_0+p_1=1$. For the composite protocol, however, this strategy allows the adversary to successfully open to any of the $2^n$ possible strings with probability 0.5. The total success probability, summed over all possible strings, is $\sum_{s \in \{0,1\}^n} p_s = 2^n \times 0.5 = 2^{n-1}$. This sum grows exponentially with $n$, thus violating any meaningful security definition for string commitment. This illustrates that a local security analysis of each component is insufficient, as it fails to account for global correlations in the adversary's strategy.

Therefore, when building cryptographic protocols from relativistic bit commitments, one must analyze security for each construction individually, since sum-binding is not composable.

\bigskip 
\section{Supplementary Note 4: Definition of ZKP}
Interactive proof systems are a powerful and versatile tool in computer science and cryptography, with applications ranging from secure communication protocols to complexity theory~\cite{Michael1988multiprover}. In a zero-knowledge interactive proof system, the prover aims to convince the verifier that a certain statement is true without revealing any additional information beyond the validity of the statement itself. This interaction occurs over multiple rounds, with the verifier challenging the prover with questions or requests for further evidence. Through this process, the verifier gains increasing confidence in the truth of the statement. Here, we formally define an interactive zero-knowledge proof system. Let $L=(L_{\text{yes}}, L_{\text{no}})$ be a language (decision problem). We formally define the three properties of interactive ZKP system. Completeness and soundness are the two properties of all interactive proof systems, and zero-knowledge is the unique property for ZKP.

\begin{definition}[Completeness]\label{completeness}

If $x\in L_{\text{yes}}$ (the statement is true), then P can convince V with high probability. Formally, if P and V follow the protocol,
\begin{equation}
\forall x \in L_{\text{yes}}, \Pr[V \text{ accepts } x] \geq 1-\varepsilon_c,  
\end{equation}
where $\varepsilon_c$ is the completeness error and it is a very small non-negative constant. Specially, when $\varepsilon_c=0$, honest P can convince V that $x\in L_{\text{yes}}$ with one hundred percent certainty, and it is called \textbf{perfect completeness}. 

\end{definition}

\begin{definition}[Soundness]\label{soundness}
    If $x\in L_{\text{no}}$ (the statement is false), then no P, even malicious P$^*$, can convince the V that 
$x\in L_{\text{yes}}$ after efficient rounds of interaction except with low probability. Formally,
\begin{equation}
    \forall x \in L_{\text{no}}, \forall \text{ malicious P's strategies }, \Pr[ \text{V accepts } x ] \le \delta_s,
\end{equation}
where $\delta_s$ is the soundness parameter and is a small positive and negligible quantity.
\end{definition}

The core property of zero-knowledge is that the verifier learns nothing except for the fact that the statement is true.  In other words, just knowing the statement (not the secret) is sufficient to imagine a scenario showing that the prover knows the secret. This is formalized by the existence of an efficient polynomial-time simulator (S) that can produce the view that looks like the real interaction between the prover and the verifier, without access to the actual secret (or witness), as shown in Supplementary Fig.~\ref{fig:zk}.

\begin{figure}[h]
	\centering
    \includegraphics[width=8.5cm]{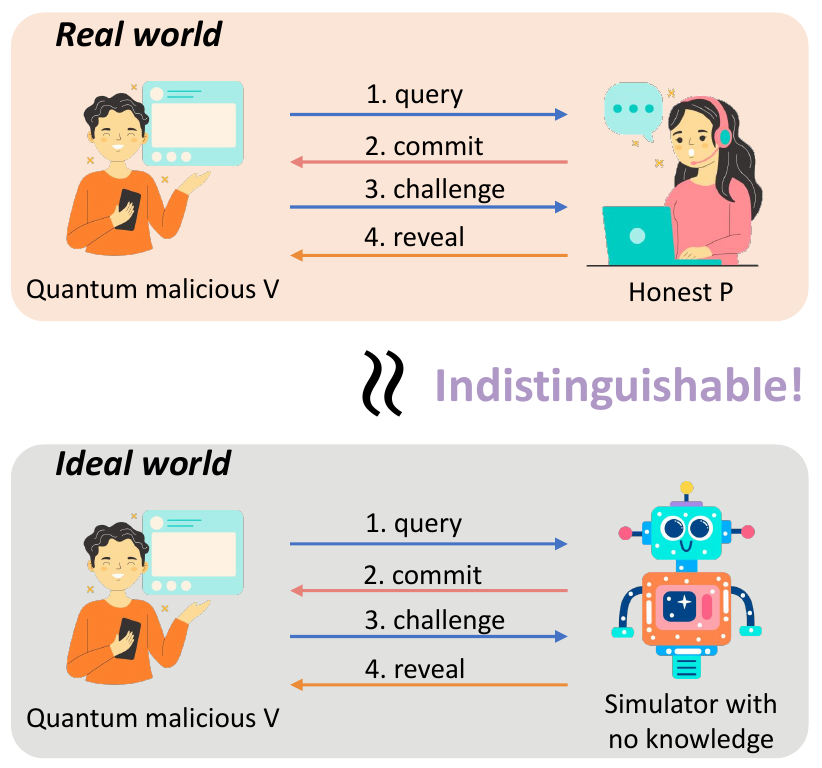}
	\caption{\textbf{Zero-knowledge.} It can be defined using a simulator with no knowledge in the ideal world, which generates a view by simulating the interaction that is indistinguishable from the interaction in the real world. The symbol $\approx$ denotes that the two probability distributions are indistinguishable.} 
\label{fig:zk}
\end{figure}

\begin{definition}[Zero-Knowledge]\label{zeroknowledge}
    An interactive proof system $(P, V)$ for a language $L$ is zero-knowledge if for any malicious probabilistic polynomial-time (PPT) verifier $V^*$, there exists a PPT simulator $S$ such that:
\begin{equation}
\forall x \in L_{\text{yes}}, \forall w \in W_x, \forall z \in \{0,1\}^*, \quad \sigma^f_{\text{real}}[P(x, w) \leftrightarrow V^*(x, z)] = \sigma^f_{\text{sim}}(x, z),
\end{equation}
where:
    \begin{itemize}
        \item $W_x$ is the set of valid witnesses for the statement $x$.
        \item $\sigma^f_{\text{real}}[P(x, w) \leftrightarrow V^*(x, z)]$ is the final view of the verifier $V^*$ during a real interaction with $P$. P possesses input statement $x$ and witness $w$. $V^*$ possesses input statement $x$ and his auxiliary input $z$.
        \item $\sigma^f_{\text{sim}}(x, z)$ is the final view simulated by $S$ on input $x$ and auxiliary input $z$.
        \item $=$ denotes perfect indistinguishability, which means the two terms are exactly identical that no distinguisher can distinguish between them even with unlimited power.
    \end{itemize}

\end{definition}
Intuitively, this definition states that the proof system $(P, V)$ is zero-knowledge if for any PPT verifier $V^*$, there exists an efficient simulator $S$ (depending on $V^*$) that can generate a view indistinguishable from a real interaction, without access to the prover's witness $w$. The auxiliary string $z$ represents prior knowledge available to $V^*$, including its random coins. The definition implies that V cannot use any prior knowledge string $z$ to mine information out of its interactions with P. Note that the definition given above is perfect zero-knowledge. Statistical (Computational) zero-knowledge is obtained by requiring that the views are statistically (computationally) indistinguishable. In this paper, we only focus on the perfect zero-knowledge, and the details of statistical and computational zero-knowledge could be found in ref.~\cite{goldreich2001foundations}.  

For the readers who are not familiar with the notions and definitions of ZKP, in the below, using an example of the graph 3-coloring problem, we give a simple explanation for the statement, witness, transcript and view to help readers better understand the definitions and proofs.
\begin{itemize}
    \item The statement is the claim (e.g., "Graph $\mathbb{G}$ is 3-colorable").
    \item The witness is the prover's secret (e.g., the actual correct 3-coloring of $\mathbb{G}$ owned by honest provers).
    \item The transcript is the complete log of the interaction between the prover and verifier.
    \item The view represents the verifier's perspective during the interactions, including all the data that the verifier can observe, such as the messages exchanged during the protocol and the randomness used in the protocol's execution.
\end{itemize}

\bigskip 
\section{Supplementary Note 5: Detailed security analysis of asymmetric relativistic ZKPs against quantum adversaries}
    
In this section, we present the security analysis of our relativistic ZKP protocol for the graph 3-coloring problem, focusing on perfect completeness and quantum perfect zero-knowledge. The proof of quantum zero-knowledge is based on the approaches in~\cite{chailloux2017relativistic,crepeau2023zeroknowledge}, which address relativistic ZKP for the Hamiltonian cycle and SAT problems, respectively.

\subsection{5.1 Proof of perfect completeness}
The definition of perfect completeness of ZKP has already been shown in Definition 2 of the main text.

\begin{theorem}
Our ZKP protocol exhibits perfect completeness: honest provers can always convince the verifiers of the graph's three-colorability.
\end{theorem}

\textit{Proof.} Theorem S3 corresponds to Theorem 2 in the main text. Let the graph $\mathbb{G}(V,E)$ be three-colorable. Honest provers assign each vertex $k \in V$ a value $y_k \in \mathbb{F}_3$, which represents the color of vertex $k$, such that for any edge $\{u,v\} \in E$, the colors of two vertices are different, i.e., $y_u\neq y_v$. In our ZKP protocol, P1 and P2 prepare a uniformly random number $b_k\in \mathbb{F}_Q$ for each vertex $k\in V$. P1 then receives $X=\{x_k\}_{k \in V}$ and calculates $A=\{a_k\}_{k \in V}$ where $a_k = x_k\cdot y_k -b_k$. Upon receiving a challenge edge $C=\{i,j\}$ from V2, P2 reveals $b_i$ and $b_j$. The verifiers can then obtain
\begin{equation}
    \begin{aligned}
        &y_i = (a_i+b_i)/x_i \\
        &y_j = (a_j+b_j)/x_j.
    \end{aligned}
\end{equation}
Since the provers are honest, $y_i \neq y_j$ holds by construction, ensuring the verification always succeeds. Thus, perfect completeness is guaranteed.

\subsection{5.2 Proof of quantum perfect zero-knowledge}
From the provers' perspective, each receives a message and responds. We assume two cheating verifiers, V1 and V2, can completely bypass timing constraints. Thus, we can consider it as a model of a single malicious verifier interacting with both provers. Furthermore, we allow the verifier to send a query to one prover after receiving a response from the other, or vice-versa. This setup is designed to model the strongest possible malicious verifier. Below, we provide a formal definition of quantum zero-knowledge, which is the general case of classical zero-knowledge where the verifier has auxiliary quantum states.

\begin{definition}[Quantum perfect zero-knowledge]
For any quantum polynomial-time verifier $V^*$ (which can deviate from the protocol and use additional quantum resources), there exists a polynomial-time simulator S such that $\forall x\in L_{\text{yes}}$, the view of verifier in the real interaction with actual provers is perfectly indistinguishable from the view simulated by the simulator S without access to the witness. Formally,
\begin{equation}
    \forall x \in L_{\text{yes}}, \forall w \in W_x, \forall z \in \{0,1\}^*, \forall \, \text{poly-qubit} \,\, \rho,\quad \sigma^f_{\text{real}}[P(x, w) \leftrightarrow V^*(x, z, \rho)] = \sigma
    ^f_{\text{sim}}(x, z, \rho),
\end{equation}

where:
    \begin{itemize}
        \item $W_x$ is the set of valid witnesses for the statement $x$.
        \item $\sigma^f_{\text{real}}[P(x, w) \leftrightarrow V^*(x, z,\rho)]$ is the final view of $V^*$ during a real interaction. P possesses input statement $x$ and witness $w$, and quantum $V^*$ possesses the auxiliary input string $z$ such as random coins and the auxiliary quantum states of polynomial qubits $\rho$.
        \item $\sigma^f_{\text{sim}}(x, z,\rho)$ is the final view simulated by $S$ on input $x$, auxiliary input $z$ and auxiliary quantum states $\rho$.
        \item $=$ denotes perfect indistinguishability, which means the two views are exactly identical that no distinguisher can distinguish between them even with unlimited power.
\end{itemize}

\end{definition}
In addition, there exist definitions of quantum statistical zero-knowledge and quantum computational zero-knowledge which could be found in~\cite{Unruh2012quantum,kobayashi2008general,ananth2021concurrent}. Here, we focus on the proof of quantum perfect zero-knowledge. 

\begin{theorem}
Our relativistic ZKP for the graph three-coloring problem is quantum perfect zero-knowledge. 
\end{theorem}

\textit{Proof.} Theorem S4 corresponds to Theorem 3 in the main text. We show that our protocol achieves quantum perfect zero-knowledge in the above model of two provers and one quantum malicious verifier. Our security analysis for quantum zero-knowledge in the relativistic setting follows the approach outlined in ref.~\cite{chailloux2017relativistic,crepeau2023zeroknowledge}. A cheating verifier is modeled as a polynomial-time uniform family of paired circuits. The verifier sends a query to P1 in a classical register $Q_1$ and a challenge to P2 in a classical register $Q_2$, receiving responses in classical registers $R_1$ and $R_2$, respectively. Additionally, the verifier has access to a private quantum register $\mathcal{V}$. To help readers better understand the proof, we give the one-to-one map between the definition notions and ZKP for the graph 3-coloring problem in Supplementary Table~\ref{tab:zkp_mapping}. 

\begin{table}[h]
\centering
\caption{Mapping between ZKP notions and the graph 3-coloring problem}
\label{tab:zkp_mapping}
\begin{tabular}{|c|c|}
\hline
\textbf{ZKP Notions} & \textbf{Graph 3-Coloring Problem} \\ \hline
Statement $ x $ & Graph $ \mathbb{G} $ is three-colorable \\ \hline
Witness $ w $ & Valid color permutations of vertices, $Y_{\pi} $ \\ \hline
Auxiliary input $ z $ & Verifier's random queries and edge, $X$ and $C$ \\ \hline
Quantum auxiliary state $ \rho $ & Verifier's poly-qubit quantum state \\ \hline
\end{tabular}
\end{table}

First of all, let us consider the view of the interaction between V$^*$ and honest P in the real world. At the beginning of the protocol, the verifier's view consists of $\sigma_0 := \rho_{\mathcal{V}}$, which is stored in his private quantum register. After the verifier's queries $X=\{x_i\}_{i\in V}$ to P1 (query phase), the verifier's view is
\begin{equation}
    \sigma^1_{\text{real}}=\sum_{X\in \mathbb{F}^{\otimes |V|}_Q} p_X\ket{X}\bra{X}_{Q_1}\otimes \rho(X)_{\mathcal{V}}.
\end{equation}
Because V$^*$ is malicious, the $X$ sent by V$^*$ is not necessarily uniformly random. Here, $p_X$ is the probability of $X$ sent to P1, and $\rho(X)_{\mathcal{V}}$ is the verifier's private quantum information after the query phase.

After the commit phase where P1 commits the answer $A:=\{a_i\}_{i\in V}$ to V$^*$, the shared classical-quantum state between the provers and the verifier is

\begin{equation}
  \sigma^2_{real}=  \frac{1}{|\Pi|}\frac{1}{Q^{|V|}}\sum_{\pi \in \Pi}\sum_{B\in \mathbb{F}_Q^{\otimes|V|}} \sum_{X\in \mathbb{F}^{\otimes |V|}_Q} p_{X}\ket{X}\bra{X}_{Q_1}\otimes\ket{A(X,Y_{\pi},B)}\bra{A(X,Y_{\pi},B)}_{R_1}\otimes \rho(X)_{\mathcal{V}}.   
\end{equation}

Here, we use entry-wise matrix multiplication and addition to represent the commitments for all vertices, where the commitments $A(X,Y_{\pi},B))=X\ast Y_{\pi}-B$, and $Y_{\pi}:=\{y_i\}_{i\in V}$, $B:=\{b_i\}_{i\in V}$. Note that $Y_{\pi}$ is determined by $\pi$ which is the random color permutation chosen by the honest provers at the beginning of this round. Since the provers are honest, variables $\pi$ and $B$ in the commit phase are uniformly random.  It is important to note that the malicious verifier can not get any useful information of coloring due to the perfect hiding of relativistic bit commitment ($B$ is uniformly random at the verifier's view).

Then in the challenge phase, malicious V$^*$ sends a challenge $C=\{i, j\}\in E$ to P2 depending on his thoughts from everything that happened before. The transcript becomes

\begin{align}
     \sigma^3_{\text{real}}= \frac{1}{|\Pi|}\frac{1}{Q^{|V|}}\sum_{\pi \in \Pi}\sum_{B\in \mathbb{F}_Q^{\otimes|V|}} \sum_{X\in \mathbb{F}^{\otimes |V|}_Q} \sum_{C\in E} &p_{X,C}\ket{X}\bra{X}_{Q_1} \otimes\ket{A(X,Y_{\pi},B)}\bra{A(X,Y_{\pi},B)}_{R_1}\otimes \ket{C}\bra{C}_{Q_2}\notag\\
     &\otimes \rho(X,C,A(X,Y_{\pi},B))_{\mathcal{V}}.
\end{align}

After the reveal phase where the verifier receives P2's final message $B(C)=\{b_i,b_j\}$ according to the challenge $C=\{i,j\}$, the final view of the real interaction becomes

\begin{align}\label{sigma_real}
   \sigma^f_{\text{real}}= \sigma^4_{\text{real}}=\frac{1}{|\Pi|}\frac{1}{Q^{|V|}}\sum_{\pi \in \Pi}\sum_{B\in \mathbb{F}_Q^{\otimes|V|}} \sum_{X\in \mathbb{F}^{\otimes |V|}_Q}\sum_{C\in E} &p_{X,C}\ket{X}\bra{X}_{Q_1}\otimes \ket{A(X,Y_{\pi},B)}\bra{A(X,Y_{\pi},B)}_{R_1}\otimes \ket{C}\bra{C}_{Q_2}\notag\\
    &\otimes \ket{B(\text{C})}\bra{B(\text{C})}_{R_2}\otimes \rho(X,C,A(X,Y_{\pi},B))_{\mathcal{V}}.
\end{align}

Now, we describe how to simulate the views without provers. We will denote the $i_{th}$ simulated view as $\sigma^i_{\text{sim}}$. Simulating $\sigma^0_{\text{real}}$ is straightforward, i.e., $\sigma^i_{\text{sim}}=\sigma^0_{\text{real}}=\rho_{\mathcal{V}}$. After the verifier's queries $X$ to the simulator (query phase), the simulated view is
\begin{equation}
    \sigma^1_{\text{sim}}=\sum_{X\in \mathbb{F}^{\otimes |V|}_Q} p_X\ket{X}\bra{X}_{Q_1}\otimes \rho(X)_{\mathcal{V}}.
\end{equation}

Then, in the commit phase, the simulator can reply with uniformly random $A^{\prime}$ as commitments because the simulator has no knowledge about the colors. Thus, the simulated view becomes
\begin{equation}
    \sigma^2_{\text{sim}}=\frac{1}{Q^{|V|}}\sum_{A^{\prime}\in \mathbb{F}_Q^{\otimes|V|}} \sum_{X\in \mathbb{F}_Q^{\otimes|V|}} p_X\ket{X}\bra{X}_{Q_1}\otimes \ket{A^{\prime}}\bra{A^{\prime}}_{R_1}\otimes \rho(X)_{\mathcal{V}}.
\end{equation}

Then in the challenge phase, V$^*$ sends $C=\{i, j\}$ to the simulator. The simulated view becomes

\begin{equation}
 \sigma^3_{\text{real}}=\frac{1}{Q^{|V|}}\sum_{A^{\prime}\in \mathbb{F}_Q^{\otimes|V|}} \sum_{X\in \mathbb{F}_Q^{\otimes|V|}}\sum_{C\in E} p_{X,C}\ket{X}\bra{X}_{Q_1}\otimes \ket{A^{\prime}}\bra{A^{\prime}}_{R_1}\otimes \ket{C}\bra{C}_{Q_2}\otimes \rho(X,C,A^{\prime})_{\mathcal{V}}.
\end{equation}

In the reveal phase, V$^*$ receives the final message $B^{\prime}(C)$ from the simulator which is selected by the simulator depending on a forgery non-valid coloring $Y^{\prime}_{\pi}$ to try to pass the verification of the verifier, so the simulated final view is
\begin{align}\label{sigma4*}
  \sigma^f_{\text{sim}}=\sigma^4_{\text{sim}}= \frac{1}{|\Pi|}\frac{1}{Q^{|V|}}\sum_{\pi \in \Pi}\sum_{A^{\prime}\in \mathbb{F}_Q^{\otimes|V|}} \sum_{X\in \mathbb{F}_Q^{\otimes|V|}}\sum_{C\in E}  &p_{X,C} \ket{X}\bra{X}_{Q_1}\otimes \ket{A^{\prime}}\bra{A^{\prime}}_{R_1}\otimes \ket{C}\bra{C}_{Q_2}\notag\\ &\otimes\ket{B^{\prime}(C)}\bra{B^{\prime}(C)}_{R_2}
 \otimes \rho(X,C,A^{\prime})_{\mathcal{V}}.
\end{align}

Below, we describe the simple simulation process and explain how $\sigma^f_{\text{real}}$ in Eq.~\ref{sigma_real} equals $\sigma^f_{\text{sim}}$ in Eq.~\ref{sigma4*}. First, the simulator sets $A^{\prime} = A(X, Y_{\pi}, B)$, which is given by:  
\begin{equation}
    A^{\prime} = A(X, Y_{\pi}, B) = X \ast Y_{\pi} - B.
\end{equation}

Since the simulator operates without relativistic constraints, it can reveal any colors for the challenge corresponding to a fixed commitment $A^{\prime} = A(X, Y_{\pi}, B)$. This allows the simulator to forge any $Y^{\prime}_{\pi}$ without knowledge of a valid three-coloring $Y_{\pi}$ and to compute a corresponding $B^{\prime}$ in polynomial time to satisfy the commitment $A(X, Y_{\pi}, B)$ from the real interaction:  
\begin{align}
    A &= X \ast Y_{\pi} - B \quad (\text{real interaction}), \\
    A^{\prime} = A &= X \ast Y^{\prime}_{\pi} - B^{\prime} \quad (\text{simulation}).
\end{align}  
Note that, the simulator should ensure $y^{\prime}_i \neq y^{\prime}_j$ in $Y^{\prime}_{\pi}$ for the challenge $C = \{i, j\}$.  

Thus, the simulator can set $A^{\prime} = A$ and replace the variable $A^{\prime}$ in Eq.~\ref{sigma4*} with $B^{\prime}$. Using the relationship $A^{\prime} = X \ast Y^{\prime}_{\pi} - B^{\prime} = A(X, Y_{\pi}, B) = X \ast Y_{\pi} - B$, we can rewrite $A(X, Y_{\pi}, B)$ as $A(X, Y^{\prime}_{\pi}, B^{\prime})$. Consequently, $\sigma^f_{\text{sim}}$ becomes:  
\begin{align}
    \sigma^f_{\text{sim}} = \frac{1}{|\Pi|} \frac{1}{Q^{|V|}} \sum_{\pi \in \Pi} \sum_{B^{\prime} \in \mathbb{F}_Q^{\otimes|V|}} \sum_{X \in \mathbb{F}_Q^{\otimes|V|}} \sum_{C \in E} &p_{X, C} \ket{X}\bra{X}_{Q_1} \otimes \ket{A(X, Y^{\prime}_{\pi}, B^{\prime})}\bra{A(X, Y^{\prime}_{\pi}, B^{\prime})}_{R_1} \notag \\ 
    &\otimes \ket{C}\bra{C}_{Q_2} \otimes \ket{B^{\prime}(C)}\bra{B^{\prime}(C)}_{R_2} \otimes \rho(X, C, A(X, Y^{\prime}_{\pi}, B^{\prime}))_{\mathcal{V}}.
\end{align}

Thus, we find that the simulator successfully simulates a view that is perfectly indistinguishable from that of the real interaction without the valid witness $Y_{\pi}$. This can be formally expressed as:  
\begin{equation}
\begin{aligned}
&\forall x \in L_{\text{yes}}, \forall \pi \in \Pi, \forall X \in \mathbb{F}_{Q}^{\otimes |V|}, \forall C \in E, \forall \text{poly-qubit} \, \rho, \\
&\sigma^f_{\text{real}}[P(x, Y_\pi) \leftrightarrow V^*(x, X, C, \rho)] = \sigma^f_{\text{sim}}(x, X, C, \rho).
\end{aligned}
\end{equation}

In conclusion, the simulation is successful, ensuring that any malicious quantum verifier gains no additional knowledge from the interactions. Therefore, our protocol achieves quantum perfect zero-knowledge.

Next, we discuss the reasons behind the simplicity of proving zero-knowledge against quantum verifiers in the relativistic setting. In this case, quantum rewinding is unnecessary, and the simulation process is straightforward. This simplicity arises from the mathematical structure of the relativistic bit commitment scheme, which allows the simulator to reveal any value for a given commitment~\cite{chailloux2017relativistic}. Specifically, even after the simulator has chosen a commitment $A^{\prime}$, it can always find a value $B^{\prime}$ that satisfies the bit commitment relation $A^{\prime} = X \cdot Y^{\prime}_{\pi} - B^{\prime}$ for any non-valid three-coloring $Y^{\prime}_{\pi}$ it wishes to reveal, where $Y^{\prime}_{\pi}$ satisfies $y^{\prime}_i \neq y^{\prime}_j$ for the challenged edge $C = \{i, j\}$.

In contrast, in classical ZKP protocols and ZK-QIPs for the graph three-coloring problem, the mathematical structure of bit commitments based on classical or quantum one-way functions does not allow the simulator to easily find values $B$ and $B^{\prime}$ such that $Y_{\pi}$ and $Y^{\prime}_{\pi}$ simultaneously satisfy the same commitment $A$ within polynomial time.  Consequently, the simulator must employ a rewinding process, where it rewinds to the initial step $\sigma_0$ to recommit the colors after knowing the challenge $C = \{i, j\}$. For a verifier with private quantum states, rewinding becomes significantly more complex because the simulator would need to replicate the initial quantum state, which is prohibited by the quantum no-cloning theorem. As a result, specialized quantum rewinding techniques are required to prove zero-knowledge against quantum verifiers~\cite{watrous2006zeroknowledge}. These ZKPs rely on computational assumptions, whereas relativistic ZKPs eliminate such dependencies by leveraging the no-signaling principle of special relativity and thus have such simple mathematical structure of bit commitment to perform such simple simulation process.

It is important to note that the simulation process described above in our relativistic ZKP cannot be employed by real provers. Real provers are subject to relativistic constraints, and the binding property of the relativistic bit commitment ensures they cannot alter previously committed colors in the real world. In the context of cryptographic proofs, the simulator's process is not a physical operation but rather a theoretical construct within an idealized model. This allows the simulator to explore execution paths until it generate a view which matches the real-world interaction within polynomial time. Because the simulator exists only as part of the mathematical framework of the proof, its seemingly 'super-powered' abilities do not compromise the protocol's soundness in the real world. For instance, in relativistic ZKPs, the simulator is not bound by relativistic constraints, and in classical ZKPs and ZK-QIPs, the simulator is permitted to perform rewinding, as these operations are confined to the idealized model.

%%%%%%%%%%%%%%%%%%%%%%% References %%%%%%%%%%%%%%%%%%%%%%%%%
% choose a style
%\bibliographystyle{ieeetr}
%\bibliographystyle{unsrt}
\bibliographystyle{naturemag}
%\bibliographystyle{apsrev}
%%%%%%%%%%%%%%%%%%%%%%%%%%%%%%%%%%%%%%%
% choose a .bib file
%\bibliography{sitext}
%%%%%%%%%%%%%%%%%%%%%%%%%%%%%%%%%%%%%%%
%\bibliographystyle{apsrev4-1}